\documentclass[]{llncs}

\usepackage{pslatex}
\usepackage{bussproofs}
\usepackage{listings}
\usepackage{fancybox}
\usepackage{mathtools}
\usepackage{tikz}
\usetikzlibrary{arrows.meta, shapes, arrows, automata, fit, matrix,
  positioning, decorations.pathreplacing, decorations.pathmorphing, calligraphy}
\usepackage{amssymb}
\usepackage{amsmath}
\usepackage{thm-restate}
\usepackage{caption}
\usepackage{subcaption}
\usepackage[hidelinks]{hyperref}
\usepackage{xcolor}
\usepackage{paralist}

\newif\ifLongVersion\LongVersiontrue

\ifLongVersion
\usepackage[createShortEnv,conf={normal}]{proof-at-the-end}
\else
\usepackage[createShortEnv,conf={end,no link to proof,restate}]{proof-at-the-end}
\fi

\ifLongVersion
\newenvironment{myLemmaE}{\begin{lemmaE}}{\end{lemmaE}}
\newenvironment{myTextE}{}{}
\else
\newenvironment{myLemmaE}{\begin{lemmaE}[][all end]}{\end{lemmaE}}
\newenvironment{myTextE}{\begin{textAtEnd}}{\end{textAtEnd}}
\fi

\renewcommand{\paragraph}[1]{\vspace*{.5\baselineskip}\noindent{\textbf{#1}}}

\newcommand{\nat}{\mathbb{N}}
\newcommand{\arityof}[1]{\#{#1}}

\newcommand{\universe}{\mathbb{U}}
\newcommand{\vars}{\mathbb{V}^{(1)}}
\newcommand{\Vars}{\mathbb{V}^{(2)}}

\newcommand{\preds}{\mathbb{A}}
\newcommand{\isdef}{\stackrel{\scriptscriptstyle{\mathsf{def}}}{=}}
\newcommand{\interv}[2]{[{#1},{#2}]}
\newcommand{\tuple}[1]{\langle {#1} \rangle}

\newcommand{\set}[1]{\{ {#1} \}}

\newcommand{\pow}[1]{\mathrm{pow}({#1})}

\newcommand{\dom}[1]{\mathrm{dom}({#1})}

\newcommand{\cardof}[1]{|\!| {#1} |\!|}

\newcommand{\pspace}{$\mathsf{PSPACE}$}

\newcommand{\signature}{\Sigma}
\newcommand{\alphabet}{\mathcal{A}}
\newcommand{\monoid}{\mathcal{M}}
\newcommand{\asucc}{\mathfrak{s}}
\newcommand{\arel}{\mathsf{R}}
\newcommand{\prel}{\mathsf{P}}
\newcommand{\acst}{\mathsf{c}}
\newcommand{\bcst}{\mathsf{b}}
\newcommand{\ecst}{\mathsf{e}}
\newcommand{\nrel}{N}
\newcommand{\ncst}{M}
\newcommand{\nrule}{R}
\newcommand{\npred}{P}

\newcommand{\ports}{\Pi}

\makeatletter
\newcommand*{\da@rightarrow}{\mathchar"0\hexnumber@\symAMSa 4B }
\newcommand*{\da@leftarrow}{\mathchar"0\hexnumber@\symAMSa 4C }
\newcommand*{\xdashrightarrow}[2][]{%
  \mathrel{%
    \mathpalette{\da@xarrow{#1}{#2}{}\da@rightarrow{\,}{}}{}%
  }%
}
\newcommand{\xdashleftarrow}[2][]{%
  \mathrel{%
    \mathpalette{\da@xarrow{#1}{#2}\da@leftarrow{}{}{\,}}{}%
  }%
}
\newcommand*{\da@xarrow}[7]{%
  \sbox0{$\ifx#7\scriptstyle\scriptscriptstyle\else\scriptstyle\fi#5#1#6\m@th$}%
  \sbox2{$\ifx#7\scriptstyle\scriptscriptstyle\else\scriptstyle\fi#5#2#6\m@th$}%
  \sbox4{$#7\dabar@\m@th$}%
  \dimen@=\wd0 %
  \ifdim\wd2 >\dimen@
    \dimen@=\wd2 %
  \fi
  \count@=2 %
  \def\da@bars{\dabar@\dabar@}%
  \@whiledim\count@\wd4<\dimen@\do{%
    \advance\count@\@ne
    \expandafter\def\expandafter\da@bars\expandafter{%
      \da@bars
      \dabar@
    }%
  }%
  \mathrel{#3}%
  \mathrel{%
    \mathop{\da@bars}\limits
    \ifx\\#1\\%
    \else
      _{\copy0}%
    \fi
    \ifx\\#2\\%
    \else
      ^{\copy2}%
    \fi
  }%
  \mathrel{#4}%
}
\makeatother

\newcommand{\store}{\nu}
\newcommand{\struc}{\sigma}
\newcommand{\strucof}[1]{\mathit{Str}({#1})}
\newcommand{\astruc}{S}
\newcommand{\adomof}[1]{D_{#1}}
\newcommand{\interpof}[1]{\struc_{#1}}
\newcommand{\glue}{\mathit{glue}}
\newcommand{\absglue}{\glue^\sharp}
\newcommand{\glueof}[2]{\glue({#1},{#2})}
\newcommand{\absglueof}[2]{\absglue({#1},{#2})}
\newcommand{\fgcst}[1]{\mathit{fgcst}_{#1}}
\newcommand{\absfgcst}[1]{\mathit{fgcst}^\sharp_{#1}}
\newcommand{\qrof}[1]{\mathrm{qr}({#1})}
\newcommand{\twsid}[1]{\asid({#1})}
\newcommand{\twformsid}[2]{\asid({#1},{#2})}
\newcommand{\typesof}[1]{\mathbb{F}^{#1}_{\scriptscriptstyle{\mathsf{MSO}}}}
\newcommand{\atype}{\tau}
\newcommand{\typeof}[2]{\mathit{type}^{#1}({#2})}
\newcommand{\absof}[1]{{#1}^\sharp}
\newcommand{\tweq}[1]{\mathit{Tw}({#1})}
\newcommand{\abstweq}[1]{\mathit{Tw}^\sharp({#1})}
\newcommand{\Dom}[1]{\mathrm{Dom}({#1})}
\newcommand{\Rel}[1]{\mathrm{Rel}({#1})}

\newcommand{\minusdom}{\mathit{rem}_\domsymb}
\newcommand{\plustd}[2]{\mathit{add}^{#1}_\domsymb({#2})}
\newcommand{\encof}[3]{\encode({#2},{#3})}

\newcommand{\comp}{\bullet}

\newcommand{\bigcomp}{\scalebox{2}{$\comp$}}

\newcommand{\predname}[1]{\mathsf{#1}}
\newcommand{\apred}{\predname{A}}
\newcommand{\bpred}{\predname{B}}

\newcommand{\emp}{\predname{emp}}

\let\Asterisk\undefined
\newcommand{\Asterisk}{\mathop{\scalebox{1.9}{\raisebox{-0.2ex}{$\ast$}}}\hspace*{1pt}}%

\newcommand{\fv}[1]{\mathrm{fv}({#1})}

\newcommand{\sol}{\textsf{SO}}
\newcommand{\mso}{\textsf{MSO}}
\newcommand{\wss}{\textsf{WS1S}}
\newcommand{\solmso}{\textsf{(M)SO}}
\newcommand{\Models}{\Vdash}

\newcommand{\seplog}{\textsf{SL}}

\newcommand{\slr}{\textsf{SLR}}

\newcommand{\asid}{\Delta}
\newcommand{\imodels}[2]{~\models^{#1}_{#2}~}

\newcommand{\arule}{\mathsf{r}}

\newcommand{\defn}[2]{\mathrm{def}_{#1}({#2})}

\newcommand{\defnof}[1]{\mathrm{def}({#1})}

\newcommand{\predtrue}{\predname{true}}

\newcommand{\lang}[1]{\mathcal{L}({#1})}
\newcommand{\langof}[2]{\mathcal{L}_{#1}({#2})}

\newcommand{\proj}[2]{{#1}\!\!\downarrow_{\scriptscriptstyle{#2}}}

\newcounter{index}

\newcommand{\vertices}{\mathcal{V}}
\newcommand{\edges}{\mathcal{E}}
\newcommand{\graph}{\mathcal{G}}
\newcommand{\strucgraph}{\struc_\graph}
\newcommand{\signagraph}{\Gamma}
\newcommand{\vertset}{\mathsf{V}}
\newcommand{\edgerel}{\mathsf{E}}
\newcommand{\tnodes}{\mathcal{N}}
\newcommand{\tedges}{\mathcal{F}}
\newcommand{\labels}{\Lambda}
\newcommand{\alabel}{\lambda}
\newcommand{\tree}{\mathcal{T}}

\newcommand{\grid}[1]{\mathcal{H}_{#1}}
\newcommand{\clique}[1]{\mathcal{K}_{#1}}
\newcommand{\iso}{\simeq}

\newcommand{\twof}[1]{\mathrm{tw}({#1})}
\newcommand{\charform}[3]{\Theta({#2},{#3})}

\newcommand{\domsymb}{\mathfrak{D}}
\newcommand{\vertex}{u}
\newcommand{\vertexSet}{U}

\newcommand{\encode}{\mathit{encode}}

\begin{document}

\title{On the Expressiveness of a Logic of Separated Relations}

\author{Radu Iosif\inst{1} \and Florian Zuleger\inst{2}}
\institute{Univ. Grenoble Alpes, CNRS, Grenoble INP, VERIMAG, 38000, France}
\institute{Institute of Logic and Computation, Technische Universit\"{a}t Wien, Austria}

\maketitle

\begin{abstract}
  We compare the model-theoretic expressiveness of the existential
  fragment of Separation Logic over unrestricted relational signatures
  (\slr) --- with only separating conjunction as logical connective and
  higher-order inductive definitions, traditionally known as the
  symbolic heap fragment --- with the expressiveness of (Monadic) Second
  Order Logic (\solmso). While \slr\ and \mso\ are incomparable on
  structures of unbounded treewidth, it turns out that \slr\ can be
  embedded in \sol, in general, and that \mso\ becomes a strict subset
  of \slr, when the treewidth of the models is bounded by a parameter
  given as input. We also discuss the problem of defining a fragment
  of \slr\ that is equivalent to \mso\ over models of bounded
  treewidth. Such a fragment would then become the most general
  Separation Logic with a decidable entailment problem, a key
  ingredient of practical verification methods for self-adapting
  (reconfigurable) component-based and distributed systems.
\end{abstract}

\section{Introduction}

Separation Logic \cite{Reynolds02,overviewSL} (\seplog) is used to
write proofs of programs that perform low-level memory updates via
pointer manipulations. The main feature of \seplog\ is the separating
conjunction connective $*$, that is associative, commutative but not
idempotent i.e., $\phi * \phi$ is not the same as $\phi$. The lack
of idempotency, traditionally recognized as a natural property of boolean
connectives, became a rather useful tool in reasoning about
\emph{resources} that can be accumulated, split or shared among the
members of a given population (see \cite{DBLP:journals/siglog/Pym19}
for a survey of resource logics and their semantics).

Since, merely a decade ago, computers were still standalone systems
(though parallel and interconnected), the most important resource of a
computing device was its embedded memory. More recently, with the
advent of complex heterogeneous systems, such as mobile networks and
cloud computing, this view is drifting towards component-based and
distributed computing, where resources are mostly understood as
components (of possibly different nature e.g., discrete and continuous
as in a cyberphysical system) and network nodes (executing processes
in disjoint memory spaces), respectively.

The modeling and verification of complex component-based and
distributed systems have been only recently addressed by logicians,
see e.g., \cite{DBLP:conf/fossacs/BolligBR19,PymVenters20}. Driven by
the needs of reasoning about such systems, we have developped dialects
of \seplog\ tailored for modeling and verification of hierarchical
component-based \cite{BozgaIosifSifakis21} and self-adapting
(dynamically reconfigurable) distributed systems
\cite{AhrensBozgaIosifKatoen21,BozgaBueriIosif22}. To this end, the
original signature of \seplog\ has been extended from a single atomic
proposition $x \mapsto (y_1, \ldots, y_k)$ interpreted as an allocated
memory cell $x$ that \emph{points-to} other non-allocated cells $y_1,
\ldots, y_k$, to a \emph{relational signature}, that contains relation
symbols (of arity greater or equal to one) and no other function
symbols than constants (function symbols of arity zero). Whereas
\seplog\ formul{\ae} are interpreted over \emph{heaps} (i.e., partial
functions with finite domain) of a fixed arity, the formul{\ae} of the
\emph{Separation Logic of Relations} (\slr) are interpreted over
generic \emph{relational structures}. In particular, the
interpretation of an atomic proposition $R(x_1, \ldots, x_n)$ is the
relation consisting of a single tuple, given by the store values of
$x_1, \ldots, x_n$, respectively.

Whereas the \seplog\ formula $x \mapsto (y_1, \ldots, y_k) * u \mapsto
(v_1, \ldots, v_k)$ constrains $x$ and $u$ to refer to distinct memory
cells (i.e., $x \neq u$), the \slr\ formula $R(x_1, \ldots, x_n) *
R(u_1, \ldots, u_n)$ means that the interpretation of the relation
symbol $R$ consists of two distinct tuples (i.e., $x_i \neq u_i$ for
at least one index $1 \leq i \leq n$). If we restrict the formul{\ae}
to contain only separation conjunctions (i.e., no boolean conjunctions
or negations), the relational dialect of \seplog\ is more general than
the original, because each points-to atom $x \mapsto (y_1, \ldots,
y_k)$ can be encoded by a formula $D(x) * H(x,y_1,\ldots,y_k)$ in a
signature with two symbols $D$ and $H$ of arities $1$ and $k+1$,
viewed as the domain $D$ and the graph $H$ of the heap,
respectively\footnote{This is not true if we allow boolean connectives
because e.g., $(x \mapsto (y) * \predtrue) \wedge (x \mapsto (y') *
\predtrue) \wedge y \neq y'$ is unsatisfiable, whereas $(D(x) * H(x,y)
* \predtrue) \wedge (D(x) * H(x,y') * \predtrue) \wedge y = y'$
remains satisfiable.}.

Despite a large body of work on the use of inductive definitions in
\seplog\ to reason about (mainly) recursive datastructures
\cite{cyclic_proofs,DBLP:conf/csl/BrotherstonFPG14,DBLP:conf/atva/EneaSW15,conf/tacas/KatelaanMZ19,DBLP:conf/lpar/KatelaanZ20,EchenimIosifPeltier21,tocl2022}
and (recently) about architectures of component-based and distributed
systems \cite{AhrensBozgaIosifKatoen21,BozgaBueriIosif22}, little is
known about the expressive power of these logics. This paper fills
this gap by pursueing a model-theoretic comparison of \slr\ with the
classical Monadic Second Order (\mso) and Second Order (\sol)
logics. We compare \slr\ with \solmso\ on structures defined over
relational signatures with and without a bound on their treewidth. The
main results (see Table \ref{tab:expressiveness} for a summary) are
that: \begin{compactenum}
\item \slr\ and \mso\ are incomparable on structures of unbounded
  treewidth i.e., there are formul{\ae} in each of the logics that do
  not have an equivalent in the other in terms of the families of
  structures they define,
\item \slr\ is strictly contained in \sol, when considering structures
  of unbounded treewidth,
\item \mso\ is strictly contained in \slr, when considering structures
  of bounded treewidth.
\end{compactenum}
The third result is probably the most interesting, as it gives a sense
of the expressive power of \slr\ (and implicitly of standard \seplog),
emphasizing the importance of the treewidth parameter.

Because the inclusion of \mso\ into \slr\ is strict over structures of
bounded treewidth, it is natural to ask for a fragment of \slr\ that
has the same expressive power as \mso, over such structures. This is
motivated by the need for a general fragment of \seplog, with a
decidable entailment problem (asking if every model of a formula
$\phi$ is also a model of another formula $\psi$, when the predicates
are interpreted by a given set of inductive definitions) useful in
designing automated program analyses. Unfortunately, this definition
is challenging because the \mso-definability of the set of models of
an \slr\ formula is undecidable, whereas the boundedness of the
treewidth for such sets remains an open problem.

\paragraph{Motivation.}
In this paragraph, we shortly sketch our earlier work that motivated the study of \slr.
Earlier versions of
\slr\ \cite{AhrensBozgaIosifKatoen21,BozgaBueriIosif22} have been used
to describe components in a component-based system (resp. network
nodes in a distributed system) by unary relation symbols and
interactions (resp. communication channels in a distributed system)
involving $n$ participants by relation symbols of arity
$n\geq2$. Moreover, recent work emphasized similarities between
recursive datastructures (e.g., lists, trees, etc.) and distributed
networks (e.g., datacenters) for instance, splay trees
\cite{SleatorTarjan85} are used to design reconfiguration algorithms
that minimize network traffic
\cite{Schmid19,DBLP:conf/infocom/PeresSGA019}. Taking this idea one
step further, we describe concisely distributed networks of unbounded
sizes, that share the same architectural style (e.g., a ring, star,
tree, etc.) in similar way as recursive datastructures are described
in \seplog\ \cite{Reynolds02}, using \emph{sets of inductive
definitions}
\cite{BozgaBueriIosif22,BozgaIosifSifakis21,AhrensBozgaIosifKatoen21}. For
instance, a system consisting of a finite but unbounded number of
components (processes) placed in a ring, such that each component
interacts only with its left and right neighbour, can be described by
the following inductive rules:
\vspace*{-.5\baselineskip}
\begin{align*}
  \mathsf{Ring}() \leftarrow & ~\exists x \exists y ~.~ I(x,y) * \mathsf{Chain}(y,x) \\
  \mathsf{Chain}(x,y) \leftarrow & ~\exists z ~.~ C(x) * I(x,z) * \mathsf{Chain}(z,y) \mid \emp * x = y
\end{align*}

\vspace*{-.5\baselineskip}\noindent
where $I(x,y)$ denotes a single interaction between nodes with
identifiers $x$ and $y$, $C(x)$ denotes a single component with
identifier $x$ and $\emp$ denotes an empty structure. The predicate
symbols $\mathsf{Ring}()$ and $\mathsf{Chain}(x,y)$ are interpreted as
the components of the pointwise smallest solution of the system of
recursive constraints above, with $\leftarrow$ interpreted as
right-to-left inclusion between sets of structures. Equivalently, any
model of $\mathsf{Ring}()$ is a model of a formula without predicate
symbols, obtained by an exhaustive unfolding of the above definitions:
\vspace*{-.4\baselineskip}
\[\mathsf{Ring}()
\Rightarrow \exists x \exists y ~.~ I(x,y) * \mathsf{Chain}(y,x)
\Rightarrow \exists x \exists y \exists z ~.~ I(x,y) * C(y) * I(y,z) *
\mathsf{Chain}(z,x) \Rightarrow ~\ldots\]

\paragraph{Related Work.}
The problem of comparing the expressiveness of
\seplog\ \cite{Reynolds02}, the most prominent among the dialects of
the calculus of Bunched Implications \cite{PymOHearn99}, with that of
classical logics received a fair amount of attention in the past. For
instance, the first proof of undecidability of the satisfiability
problem for first-order \seplog\ \cite{DBLP:conf/fsttcs/CalcagnoYO01}
is based on a reduction to Trakhtenbrot's undecidability result for
first-order logic on finite models
\cite{DBLP:books/daglib/0082516}. This proof was reviewed with certain
criticism, given that the only construction specific to \seplog\ is
the intuitionistic binary points-to formula used to encode arbitrary
binary relations as $R(x,y) \isdef \exists z ~.~ z \hookrightarrow
(x,y)$, where $z \hookrightarrow (x,y)$ stands for $z \mapsto (x,y) *
\predtrue$. A more refined proof of undecidability for unary points-to
relations $x \mapsto y$ was given by Brochenin, Demri and Lozes
\cite{DBLP:journals/iandc/BrocheninDL12}. In particular, they show
that \sol\ over heaps with one record field has the same expressivity
as \seplog\ with unary points-to predicates.

A related line of work, pioneered by Lozes~\cite{PhD-lozes}, is the
translation of quantifier-free \seplog\ formul{\ae} into boolean
combinations of \emph{core formul{\ae}}, belonging to a small set of
very simple patterns. This enables a straightforward translation of
the quantifier-free fragment of \seplog\ into first-order logic, over
unrestricted signatures with both relation and function symbols, that
was subsequently extended to two quantified variables
\cite{DBLP:journals/tocl/DemriD16} and restricted quantifier prefixes
\cite{DBLP:journals/tocl/EchenimIP20}. Moreover, a translation of
quantifier-free \seplog\ into first-order logic, based on the small
model property of the former, has been described in
\cite{CalcagnoGardnerHague05}.

Unlike our work, the above references focus exclusively on fragments
of \seplog\ without inductively defined predicates, but with arbitrary
combinations of boolean (conjunction, implication) and multiplicative
(separating conjunction, magic wand) connectives. A non-trivial
attempt of generalizing the technique of translating \seplog\ into
boolean combinations of core formul{\ae} to \emph{reachability} and
\emph{list segment} predicates is given in
\cite{DBLP:journals/tocl/DemriLM21}. Moreover, an in-depth comparison
between the expressiveness of various models of separation i.e.,
spatial, as in \seplog, and contextual, as in ambient logics, can be
found in \cite{DBLP:phd/hal/Mansutti20}.

To the best of our knowledge, the results reported in this paper
constitute the first model-theoretic comparison of a Separation Logic
with \emph{arbitrary inductive predicates}, written using a
generalized relational signature and interpreted over logical
structures (instead of heaps) with the \solmso\ logics. 
Previous work
along this line concerns a fairly general fragment of \seplog, defined
using syntactic restrictions on the rules used to interpret the
predicates, that translates into \mso\ forml{\ae} and has models of
bounded treewidth \cite{DBLP:conf/cade/IosifRS13}. 
This fragment has also been studied in~\cite{conf/tacas/KatelaanMZ19,tocl2022}, where an alternative decision procedure based on types has been developed.
Our result showing the embedding of \mso\ into \slr, on models of bounded treewidth, based on the refinement of a generic SID for bounded treewidth structures by \mso\-types, is dual and opens the (hard) problem of defining a fragment of \slr\ with the same expressivity as \mso.
The approach of refining SIDs has previously been exploited for the study of robustness properties~\cite{conf/esop/JansenKMNZ17} but not for the comparison of \seplog\ and \mso.

\section{Definitions}
\label{sec:defs}

For a set $A$, we denote by $\pow{A}$ its powerset and $A^1 \isdef A$,
$A^{i+1} \isdef A^i \times A$, for all $i \geq 1$, where $\times$ is
the Cartesian product, and $A^+ \isdef \bigcup_{i\geq1} A^i$. The
cardinality of a finite set $A$ is denoted by $\cardof{A}$. Given
integers $i$ and $j$, we write $\interv{i}{j}$ for the set
$\set{i,i+1,\ldots,j}$, assumed to be empty if $i>j$.

\paragraph{Signatures and Structures}
Let $\signature = \set{\arel_1, \ldots, \arel_\nrel, \acst_1, \ldots,
  \acst_\ncst}$ be a finite \emph{signature}, where $\arel_i$ denote
relation symbols of arity $\arityof{\arel_i}$ and $\acst_j$ denote
constant symbols (i.e., nullary function symbols). Since there are no
function symbols of arity more than zero, these signatures are said to
be \emph{relational} in the literature, see e.g.,
\cite{courcelle_engelfriet_2012}.

A \emph{structure} is a pair $(\universe, \struc)$, where $\universe$
is an infinite countable set, called \emph{universe}, and $\struc :
\signature \rightarrow \universe \cup \pow{\universe^+}$ is a function
that maps each relation symbol $\arel_i$ to a relation
$\struc(\arel_i) \subseteq \universe^{\arityof{\arel_i}}$ and each
constant $\acst_j$ to an element of the universe. We define the sets
$\Rel{\struc} \isdef \set{u_k \mid \tuple{u_1, \ldots,
    u_{\arityof{\arel_i}}} \in \struc(\arel_i),~ i \in
  \interv{1}{\nrel},~ k \in \interv{1}{\arityof{\arel_i}}}$ and
$\Dom{\struc} \isdef \Rel{\struc} \cup \set{\struc(\acst_1), \ldots,
  \struc(\acst_\ncst)}$ of elements that belong to the interpretation
of a relation symbols or a constant from $\signature$, respectively.
Unless stated otherwise (as in e.g., \S\ref{sec:k-mso-slr}), we consider
the signature $\signature$ and the countably infinite universe
$\universe$ to be fixed, for all structures considered in the rest of
this paper. Hence we denote structures simply by $\struc$, in the
following.

\begin{remark}\label{rem:structures}
  The standard notion of structure from the literature encapsulates
  the universe (sometimes called \emph{domain}), such that two
  different structures may have different universes (see e.g.,
  \cite{DBLP:books/daglib/0082516} for the standard definitions). By
  considering that all structures share the same universe we deviate
  from the standard definition. This is important for the definition
  of a composition operation on structures that extends naturally the
  union of heaps with disjoint domains, defined over the same set of
  memory locations, from the semantics of
  \seplog\ \cite{Reynolds02}. Moreover, standard structures may have a
  finite universe, whereas here we consider a countably infinite
  universe, in analogy with the set of memory
  locations. \hfill$\blacksquare$
\end{remark}

\begin{definition}\label{def:composition}
  Two structures $\struc_1$ and $\struc_2$ are \emph{disjoint} if and
  only if $\struc_1(\arel_i) \cap \struc_2(\arel_i) = \emptyset$, for
  all $i \in \interv{1}{\nrel}$ and \emph{compatible} if and only if
  $\struc_1(\acst_j) = \struc_2(\acst_j)$, for all $j \in
  \interv{1}{\ncst}$.  The \emph{composition} of two disjoint and
  compatible structures is the structure $\struc_1 \comp \struc_2$,
  such that $(\struc_1 \comp \struc_2)(\arel_i) = \struc_1(\arel_i)
  \cup \struc_2(\arel_i)$ and $(\struc_1 \comp \struc_2)(\acst_j) =
  \struc_1(\acst_j) = \struc_2(\acst_j)$, for all $i \in
  \interv{1}{\nrel}$ and $j \in \interv{1}{\ncst}$. The composition is
  undefined for structures that are not disjoint or not compatible.
\end{definition}

In the rest of this paper, we will identify structures that differ by
a renaming of elements from the universe and consider only classes of
structures that are closed under isomorphism:

\begin{definition}\label{def:isomorphism}
  Two structures $\struc_1$ and $\struc_2$ are \emph{isomorphic},
  denoted $\struc_1 \iso \struc_2$ if and only if there exists a
  bijection $h : \universe \rightarrow \universe$, such that: \begin{compactenum}
  \item for each $i \in \interv{1}{\nrel}$ and $\tuple{u_1, \ldots,
    u_{\arityof{\arel_i}}} \in \universe^{\arityof{\arel_i}}$, we have
    $\tuple{u_1, \ldots, u_{\arityof{\arel_i}}} \in \struc_1(\arel)
    \iff \tuple{h(u_1), \ldots, h(u_{\arityof{\arel_i}})} \in
    \struc_2(\arel)$,
  \item for each $j \in \interv{1}{\ncst}$, we have $h(\struc_1(c_j)) = \struc_2(c_j)$.
  \end{compactenum}
\end{definition}

\vspace*{-.5\baselineskip}
\ifLongVersion
\paragraph{Graphs, Cliques, Trees and Grids}
\fi
A graph is a pair $\graph = (\vertices,\edges)$, such that $\vertices
\subseteq \universe$ is a finite set of \emph{vertices} and $\edges
\subseteq \vertices \times \vertices$ is a set of \emph{edges}. All
graphs considered in this paper are directed i.e., $\edges$ is not a
symmetric relation. The following definition relates graphs and
structures.

\begin{definition}\label{def:graph-struc}
  Each graph $\graph = (\vertices,\edges)$ can be viewed as a
  structure $\strucgraph$ over the signature $\signagraph \isdef
  \set{\vertset, \edgerel}$, where $\arityof{\vertset}=1$ and
  $\arityof{\edgerel}=2$, such that $\strucgraph(\vertset) =
  \vertices$ and $\strucgraph(\edges) = \edgerel$.
\end{definition}
A \emph{path} in $\graph$ is a sequence of pairwise distinct vertices
$v_1, \dots, v_n \in \vertices$, such that $(v_i,v_{i+1}) \in \edges$
or $(v_{i+1},v_i) \in \edges$, for all $i \in \interv{1}{n-1}$. A set
of vertices $V \subseteq \vertices$ is \emph{connected in $\graph$} if
and only if there is path in $\graph$ between any two vertices in
$V$. A graph $\graph$ is \emph{connected} if and only if $\vertices$
is connected in $\graph$. A \emph{clique} is a graph such that there
exists an edge between each two nodes. We denote by $\clique{n}$ the
set of (pairwise isomorphic) cliques with $n$ nodes.

A \emph{tree} is a tuple $\tree = (\tnodes,\tedges,r,\alabel)$, where
$(\tnodes,\tedges)$ is a graph, $r \in \tnodes$ is a designated vertex
called the \emph{root}, such that there exists a path in
$(\tnodes,\tedges)$ from $r$ to any other vertex $v \in \tnodes$, $r$
has no incoming edge $(p,r) \in \tedges$ and no vertex $n \in \tnodes$
has two incoming edges $(m,n), (p,n) \in \tedges$, for $m\neq p$. The
mapping $\alabel : \tnodes \rightarrow \labels$ associates each vertex
of the tree with a \emph{label} from a given set $\labels$.

\ifLongVersion A $n \times m$ \emph{grid} is a graph
$(\vertices,\edges)$, such that there exists a bijective function $g :
\vertices \rightarrow \interv{1}{n} \times \interv{1}{m}$ and, for
each edge $(u,v) \in \edges$, such that $g(u) = (i,j)$, either $i < n$
and $g(v) = (i+1,j)$, or $j < m$ and $g(v) = (i,j+1)$. A grid is
\emph{square} if and only if $n=m$, and we denote by $\grid{n}$ the
set of (pairwise isomorphic) $n \times n$ square grids.  \fi

\paragraph{Treewidth}
We define a measure on structures indicating how close a structure
is to a tree:

\begin{definition}\label{def:treewidth}
  A \emph{tree decomposition} of a structure $\struc$ over the
  signature $\signature$ is a tree $\tree=(\tnodes,\tedges,r,\alabel)$
  labeled with subsets of $\universe$, such that the following
  hold: \begin{compactenum}
  \item\label{it1:treewidth} for each relation symbol $\arel \in
    \signature$ and each tuple $\tuple{u_1, \ldots,
      u_{\arityof{\arel}}} \in \struc(\arel)$ there exists $n \in
    \tnodes$, such that $\set{u_1, \ldots, u_{\arityof{\arel}}}
    \subseteq \alabel(n)$, and
  \item\label{it2:treewidth} for each $u \in \Dom{\struc}$, the set
    $\set{n \in \tnodes \mid u \in \alabel(n)}$ is non-empty and
    connected in $(\tnodes,\tedges)$.
    %
  \end{compactenum}
  The \emph{width} of the tree decomposition is $\twof{\tree} \isdef
  \max_{n \in \tnodes} \cardof{\alabel(n)}-1$. The \emph{treewidth} of
  the structure $\struc$ is $\twof{\struc} \isdef \min
  \set{\twof{\tree} \mid \tree \text{ is a tree decomposition of }
    \struc}$.
\end{definition}

It is clear that two isomorphic structures have the same treewidth.
In the following, we consider indexed families of structures that are
closed under isomorphism. Such a family is \emph{treewidth-bounded} if
and only if the set of corresponding treewidths is finite and
\emph{treewidth-unbounded} otherwise. A family is \emph{strictly
treewidth-unbounded} if and only if it is treewidth-unbounded and any
infinite subfamily is treewidth-unbounded. The following result can be
found in several textbooks (see e.g.,
\cite{courcelle_engelfriet_2012}) and is restated here for
self-containment:

\ifLongVersion
\begin{propositionE}\label{prop:strictly-unbounded}
  The indexed families
  $\set{\clique{n} \mid n \in \nat}$ and
  $\set{\grid{n} \mid n \in \nat}$
  are strictly treewidth-unbounded.
\end{propositionE}
\begin{proofE}
  \noindent1. Let $\clique{n} \isdef
  (\vertices, \edges)$, modulo isomorphism. We prove that
  $\twof{\clique{n}}=n-1$, for all $n \geq 2$, by showing that, in
  every tree decomposition $(\tnodes,\tedges,r,\alabel)$ of
  $\clique{n}\isdef(\vertices,\edges)$ there is a node $p\in\tnodes$,
  such that $\alabel(p)=\vertices$. By induction on $n\geq2$, the base
  case $n=2$ follows immediately from point (\ref{it1:treewidth}) of
  Def. \ref{def:treewidth}. For the inductive step $n > 2$, let $u \in
  \vertices$ be a vertex and let $p\in\tnodes$ be the node of least
  depth, such that $u \in \alabel(p)$. By point (\ref{it2:treewidth})
  of Def. \ref{def:treewidth}, this node is unique. Let $\set{v_1,
    \ldots, v_{n-1}} \isdef \vertices\setminus\set{u}$. By the
  inductive hypothesis, there exists a node $m \in \tnodes$, such that
  $\alabel(m)=\set{v_1, \ldots, v_{n-1}}$, because the restriction of
  $(\vertices,\edges)$ to $\set{v_1, \ldots, v_{n-1}}$ is a clique and
  $(\tnodes,\tedges,r,\alabel)$ is also a tree decomposition of that
  clique. Since $(u,v_1), \ldots, (u,v_{n-1})\in\edges$, by point
  (\ref{it1:treewidth}) of Def. \ref{def:treewidth}, there exists
  nodes $q_1, \ldots, q_{n-1} \in \tnodes$, such that $u,v_i \in
  \alabel(q_i)$, for all $i \in \interv{1}{n-1}$. By the choice of
  $p$, these nodes are all descendants of $p$, because $u \in
  \alabel(q_i)$, for all $i \in \interv{1}{n-1}$ and $p$ has the
  lowest depth among all such nodes. We distinguish two
  cases: \begin{compactitem}
  \item $m$ is not a descendant of $p$, then let $q$ be the deepest
    common ancestor of $m$ and $p$. Since $v_i \in \alabel(q_i) \cap
    \alabel(m)$, we obtain $v_i \in \alabel(q)$, for all $i \in
    \interv{1}{n-1}$, by point (\ref{it2:treewidth}) of
    Def. \ref{def:treewidth}. This leads to $u, v_1, \ldots, v_{n-1}
    \in \alabel(p)$, by the same argument.
  \item $m$ is a descendant of $p$ and let $q'_i, \ldots, q'_{n-1}$ be
    the deepest common ancestors of $m$ and $q_1, \ldots, q_{n-1}$,
    respectively. Since $v_i \in \alabel(q_i) \cap \alabel(m)$, we
    obtain $v_i \in \alabel(q'_i)$, for all $i \in \interv{1}{n-1}$,
    by point (\ref{it2:treewidth}) of Def. \ref{def:treewidth}. Since
    $q'_1, \ldots, q'_{n-1}$ are all ancestors of $m$, they are
    linearly ordered by the ancestor relation and let $q'_k$ be the
    deepest such node. Then we obtain $u, v_1, \ldots, v_{n-1} \in
    \alabel(q'_k)$, by point (\ref{it2:treewidth}) of
    Def. \ref{def:treewidth}.
  \end{compactitem}

  \vspace*{\baselineskip}\noindent2. Let
  $\grid{n}\isdef(\vertices,\edges)$. Consider the $k$-cops and robber
  game on $(\vertices,\edges)$, defined as follows. A position in the
  game is a pair $(C,r)$, where $C \subseteq \vertices$,
  $\cardof{C}=k$ and $r \in \vertices \setminus C$. The game can move
  from $(C_i,r_i)$ to $(C_{i+1},r_{i+1})$ if there exists a path
  between $r_i$ and $r_{i+1}$ in the restriction of
  $(\vertices,\edges)$ to $\vertices \setminus C_i\cap C_{i+1}$. We
  say that $k$ cops catch the robber if and only if every sequence of
  moves in the game is finite. It is known that, if $\twof{\graph}\leq
  k$ then $k+1$ cops catch the robber on a graph $\graph$
  \cite{SEYMOUR199322}. Since $n-1$ cops do not catch the robber
  (which can always move to the intersection of a cop-free row and a
  cop-free column) it follows that $\twof{\grid{n}} \geq n-1$.
\end{proofE}
\fi

\section{Logics}
\label{sec:logics}

We introduce two logics for reasoning about structures over a (fixed)
relational signature $\signature = \set{\arel_1, \ldots, \arel_\nrel,
  \acst_1, \ldots, \acst_\ncst}$. The first such logic, called the
\emph{Separation Logic of Relations} (\slr), uses a set of
\emph{first-order variables} $\vars = \set{x,y,\ldots}$ and a set of
\emph{predicates} $\preds = \set{\apred, \bpred, \ldots}$ of arities
$\arityof{\apred}$, $\arityof{\bpred}$, etc. We use the symbols $\xi,
\chi \in \vars \cup \set{\acst_1, \ldots, \acst_\ncst}$ to denote
\emph{terms} i.e., either first-order variables or constants. The
formul{\ae} of \slr\ are defined by the syntax:
\vspace*{-.5\baselineskip}
\[\phi := \emp \mid \xi=\chi \mid \xi\neq\chi \mid \arel(\xi_1, \ldots, \xi_{\arityof{\arel}}) \mid
\apred(\xi_1, \ldots, \xi_{\arityof{\apred}}) \mid \phi * \phi \mid
\exists x ~.~ \phi\]

\vspace*{-.5\baselineskip}\noindent The formul{\ae} $\xi=\chi$ and
$\xi\neq\chi$ are called \emph{equalities} and \emph{disequalities},
$\arel(\xi_1, \ldots,\xi_{\arityof{\arel}})$ and
$\apred(\xi_1,\ldots,\xi_{\arityof{\apred}})$ are called
\emph{relation} and \emph{predicate atoms}, respectively. By
\emph{atom} we mean either $\emp$ or any of the above atomic
formul{\ae}.

A formula with no occurrences of predicate atoms (resp. existential
quantifiers) is called \emph{predicate-free}
(resp. \emph{quantifier-free}). A variable is \emph{free} if it does
not occur within the scope of an existential quantifier and let
$\fv{\phi}$ be the set of free variables of $\phi$. A \emph{sentence}
is a formula with no free variables. A \emph{substitution}
$\phi[x_1/\xi_1 \ldots x_n/\xi_n]$ replaces simultaneously every
occurrence of the free variable $x_i$ by the term $\xi_i$ in $\phi$,
for all $i \in \interv{1}{n}$. Before defining the semantics of
\slr\ formul{\ae}, we consider definitions that
assign meaning to predicates:

\begin{definition}\label{def:sid}
  A \emph{set of inductive definitions (SID)} $\asid$ consists of
  \emph{rules} $\apred(x_1, \ldots, x_{\arityof{\apred}}) \leftarrow
  \phi$, where $x_1, \ldots, x_{\arityof{\apred}}$ are pairwise
  distinct variables, called \emph{parameters}, such that $\fv{\phi}
  \subseteq \set{x_1, \ldots, x_{\arityof{\apred}}}$. An atom $\alpha$
  \emph{occurs} in a rule $\apred(x_1, \ldots, x_{\arityof{\apred}})
  \leftarrow \phi$ if and only if it occurs in $\phi$. We say that
  the rule $\apred(x_1, \ldots, x_{\arityof{\apred}}) \leftarrow \phi$
  \emph{defines} $\apred$. We denote by $\defn{\asid}{\apred}$ the set
  of rules from $\asid$ that define $\apred$ and by $\defnof{\asid}$
  the set of predicates defined by some rule in $\asid$.
\end{definition}
Note that having distinct parameters in a rule is without loss of
generality, as e.g., a rule $\apred(x_1, x_1) \leftarrow \phi$ can be
equivalently written as $\apred(x_1, x_2) \leftarrow x_1 = x_2 *
\phi$. As a convention, we shall always use the names $x_1, \ldots,
x_{\arityof{\apred}}$ for the parameters of a rule that defines
$\apred$.

The semantics of \slr\ formul{\ae} is given by the satisfaction relation
$(\struc,\store) \models_\asid \phi$ between structures and
formul{\ae}. This relation is parameterized by a \emph{store} $\store
: \vars \rightarrow \universe$ mapping the free variables of a formula
into elements of the universe and an SID $\asid$. We write $\store[x
  \leftarrow u]$ for the store that maps $x$ into $u$ and agrees with
$\store$ on all variables other than $x$. For a term $\xi$, we denote
by $(\struc,\store)(\xi)$ the value $\struc(\xi)$ if $\xi$ is a
constant, or $\store(\xi)$ if $\xi$ is a first-order variable. The
satisfaction relation is defined by induction on the structure of
formul{\ae}, as follows:
\vspace*{-.2\baselineskip}
\[\begin{array}{rclcl}
(\struc,\store) & \models_\asid & \emp & \iff &
\struc(\arel) = \emptyset \text{, for all } \arel\in\signature \\
(\struc,\store) & \models_\asid & \xi \bowtie \chi & \iff &
(\struc,\store) \models \emp \text{ and } (\struc,\store)(\xi) \bowtie (\struc,\store)(\chi) \text{, for all } \bowtie\in\!\set{=,\neq} \\
(\struc,\store) & \models_\asid & \arel(\xi_1, \ldots, \xi_{\arityof{\arel}}) & \iff &
\struc(\arel) = \set{\tuple{(\struc,\store)(\xi_1), \ldots, (\struc,\store)(\xi_{\arityof{\arel}})}} \text{ and } \struc(\arel') = \emptyset, \\
&&&& \text{for all } \arel' \in \signature \setminus\set{\arel} \\
(\struc,\store) & \models_\asid & \apred(\xi_1, \ldots, \xi_{\arityof{\apred}}) & \iff &
(\struc,\store) \models_\asid \phi[x_1/\xi_1, \ldots, x_{\arityof{\apred}}/\xi_{\arityof{\apred}}]
\text{, for some rule } \\ &&&& \apred(x_1, \ldots, x_{\arityof{\apred}}) \leftarrow \phi \text{ from } \asid \\
(\struc,\store) & \models_\asid & \phi_1 * \phi_2 & \iff &
\text{there exist structures } \struc_1 \text{ and } \struc_2 \text{, such that } \\
&&&&\struc = \struc_1 \comp \struc_2 \text{ and } (\struc_i,\store) \models_\asid \phi_i \text{, for all } i = 1,2 \\
(\struc,\store) & \models_\asid & \exists x ~.~ \phi & \iff &
(\struc,\store[x\leftarrow u]) \models_\asid \phi \text{, for some } u \in \universe
\end{array}\]

\vspace*{-.2\baselineskip}\noindent If $\phi$ is a sentence, the
satisfaction relation $(\struc,\store) \models_\asid \phi$ does not
depend on the store, written $\struc \models_\asid \phi$, in which
case we say that $\struc$ is a \emph{model} of $\phi$. If $\phi$ is a
predicate-free formula, the satisfaction relation does not depend on
the SID, written $(\struc,\store) \models \phi$.  We say that a pair
$(\phi,\asid)$, consisting of a sentence $\phi$ and an SID $\asid$,
\slr-\emph{defines} a set $\mathcal{S}$ of structures if and only if
$\struc \models_\asid \phi \iff \struc \in \mathcal{S}$. By the result
below, every set of \slr-defined structures is a union of equivalence
classes of isomorphism (Def. \ref{def:isomorphism}):

\begin{propositionE}\label{prop:sl-iso}
  Given structures $\struc \iso \struc'$, for any sentence
  $\phi$ of \slr\ and any SID $\asid$, we have $\struc \models_\asid
  \phi \iff \struc' \models_\asid \phi$.
\end{propositionE}
\begin{proofE}
  By induction on the definition of the satisfaction relation
  $\models_\asid$, we show that $(\struc,\store) \models_\asid \psi
  \iff (\struc',h \circ \store) \models_\asid \psi$, for any store
  $\store$ and any bijection $h : \universe \rightarrow \universe$,
  such that, for any relation symbol $\arel \in \signature$, we have
  $\tuple{u_1, \ldots, u_n} \in \struc(\arel) \iff \tuple{h(u_1),
    \ldots, h(u_n)} \in \struc'(\arel)$ and, for any constant $\acst
  \in \signature$, we have $h(\struc(\acst))=\struc'(\acst)$, as in
  Def. \ref{def:isomorphism}. We consider the following
  cases: \begin{compactitem}
  \item $\phi = \emp$: for all $\arel\in\signature$, we have
    $\struc(\arel)=\emptyset \iff \struc'(\arel)=\emptyset$, by the
    existence of $h$.
  \item $\phi = \xi \bowtie \chi$, for $\bowtie \in \set{=,\neq}$: we prove only
    the case where $\xi\in\vars$ and
    $\chi\in\set{\acst_1,\ldots,\acst_\ncst}$, the other cases being
    similar. By the above point, for all $\bowtie \in \set{=,\neq}$, we
    have $\struc \models \emp \iff \struc' \models \emp$ and
    $\store(\xi) \bowtie \struc(\chi) \iff h(\store(\xi)) \bowtie
    h(\struc(\chi)) \iff (h \circ \store)(\xi) \bowtie \struc'(\chi)$.
  \item $\phi = \arel(\xi_1, \ldots, \xi_{\arityof{\arel}})$:
    $\tuple{(\struc,\store)(\xi_1), \ldots,
    (\struc,\store)(\xi_{\arityof{\arel}})} \in \struc(\arel) \iff
    \tuple{h((\struc,\store)(\xi_1)), \ldots,
      h((\struc,\store)(\xi_{\arityof{\arel}}))} \in
    \struc'(\arel)$. If $\xi_1, \ldots, \xi_{\arityof{\arel}} \in
    \vars$, the latter condition is $\tuple{(h\circ\store)(\xi_1),
      \ldots, (h\circ\store)(\xi_{\arityof{\arel}})} \in
    \struc'(\arel)$. Else, if $\xi_1, \ldots, \xi_{\arityof{\arel}}
    \in \set{\acst_1, \ldots, \acst_\ncst}$, the condition is
    $\tuple{h(\struc(\xi_1)), \ldots,
      h(\struc(\xi_{\arityof{\arel}}))} \in \struc'(\arel)$. The
    general case $\xi_1, \ldots, \xi_{\arityof{\arel}} \in \vars \cup
    \set{\acst_1, \ldots, \acst_\ncst}$ is a combination of the above
    cases.
  \item $\phi = \apred(\xi_1, \ldots, \xi_{\arityof{\apred}})$: this
    case follows by the induction hypothesis.
  \item $\phi = \phi_1 * \phi_2$: $(\struc,\store) \models_\asid
    \phi_1 * \phi_2$ if and only if there exists disjoint and
    compatible structures $\struc_1 \comp \struc_2 = \struc$, such
    that $(\struc_i,\store) \models_\asid \phi_i$, for all $i =
    1,2$. We define the structures $\struc'_1$ and $\struc'_2$ as
    follows, for $i = 1,2$:: \begin{compactitem}
    \item $\struc'_i(\arel) = \set{\tuple{h(u_1), \ldots,
        h(u_{\arityof{\arel}})} \mid \tuple{u_1, \ldots,
        u_{\arityof{\arel}}} \in \struc_i(\arel)}$, for all relation
      symbols $\arel \in \signature$,
    \item $\struc'_i(\acst) = h(\struc_i(\acst))$, for all constants
      $\acst \in \signature$.
    \end{compactitem}
    Then $\struc'_1$ and $\struc'_2$ are disjoint and compatible and
    $\struc' = \struc'_1 \comp \struc'_2$. Moreover, by the inductive
    hypothesis, we have $(\struc'_i,\store) \models_\asid \phi_i$, for
    all $i = 1,2$, leading to $(\struc',\store) \models_\asid \phi_1 *
    \phi_2$.
  \item $\phi = \exists x ~.~ \psi$: by the inductive hypothesis, we
    obtain $(\struc,\store[x\leftarrow u]) \models_\asid \psi \iff
    (\struc',h\circ(\store[x\leftarrow u])) \models_\asid \psi \iff
    (\struc',(h\circ\store)[x \leftarrow h(u)]) \models_\asid \psi
    \iff (\struc',h\circ\store)\models_\asid \exists x ~.~ \psi$.
  \end{compactitem}
\end{proofE}

The other logic is \emph{Second Order Logic} (\sol) defined using a
set of \emph{second-order variables} $\Vars = \set{X,Y,\ldots}$, in
addition to first-order variables. As usual, we denote by
$\arityof{X}$ the arity of a second-order variable $X$. Terms and
atoms are defined in the same way as in \slr. The formul{\ae} of
\sol\ are defined by the following syntax:
\vspace*{-.5\baselineskip}
\[\psi := \xi=\chi \mid \arel(\xi_1, \ldots, \xi_{\arityof{\arel}}) \mid X(\xi_1, \ldots, \xi_{\arityof{X}})
\mid \neg\psi \mid \psi \wedge \psi \mid \exists x ~.~ \psi \mid
\exists X ~.~ \psi\]

\vspace*{-.5\baselineskip}\noindent
As usual, we write $\xi \neq \chi \isdef \neg \xi = \chi$,
$\psi_1 \vee \psi_2 \isdef \neg(\neg\psi_1 \wedge \neg\psi_2)$,
$\psi_1 \rightarrow \psi_2 \isdef \neg\psi_1 \vee \psi_2$, $\forall x
~.~ \psi \isdef \neg\exists x ~.~ \neg\psi$ and $\forall X ~.~ \psi
\isdef \neg\exists X ~.~ \neg\psi$. We denote by \mso\ the fragment of
\sol\ restricted to using only second-order variables of arity one.

The semantics of \sol\ is given by the relation $(\struc,\store)
\Models \psi$, where the store $\store : \vars \cup \Vars \rightarrow
\universe \cup \bigcup_{k\geq1}\pow{\universe^k}$ maps each
first-order variable $x \in \vars$ to an element of the universe
$\store(x) \in \universe$ and each second-order variable $X \in \Vars$
to a relation $\store(X) \subseteq \universe^{\arityof{X}}$. The
satisfaction relation is defined inductively on the structure of
formul{\ae}
\vspace*{-.5\baselineskip}
\ifLongVersion
:
\[\begin{array}{rclcl}
(\struc,\store) & \Models & \xi=\chi & \iff & (\struc,\store)(\xi)=(\struc,\store)(\chi) \\
(\struc,\store) & \Models & \arel(\xi_1, \ldots, \xi_{\arityof{\arel}}) & \iff &
\tuple{(\struc,\store)(\xi_1), \ldots, (\struc,\store)(\xi_{\arityof{\arel}})} \in \struc(\arel) \\
(\struc,\store) & \Models & X(\xi_1, \ldots, \xi_{\arityof{X}}) & \iff & \tuple{(\struc,\store)(\xi_1), \ldots, (\struc,\store)(\xi_{\arityof{X}})} \in \store(X) \\
(\struc,\store) & \Models & \neg\psi & \iff & (\struc,\store) \not\Models \psi \\
(\struc,\store) & \Models & \psi_1 \wedge \psi_2 & \iff & (\struc,\store) \Models \psi_i \text{, for all } i =1,2 \\
(\struc,\store) & \Models & \exists x ~.~ \psi & \iff & (\struc,\store[x\leftarrow u]) \Models \psi \text{, for some } u \in \universe \\
(\struc,\store) & \Models & \exists X ~.~ \psi & \iff & (\struc,\store[X\leftarrow U]) \Models \psi
\text{, for some } U \subseteq \universe^{\arityof{X}}
\end{array}\]

\vspace*{-.5\baselineskip}\noindent
\fi We write $\struc \Models \phi$
whenever $\phi$ is a sentence. A sentence $\phi$ \sol-\emph{defines} a
set of structures $\mathcal{S}$ if and only if $\struc \Models \phi
\iff \struc \in \mathcal{S}$. It is well-known (see e.g.,
\cite{DBLP:books/daglib/0082516}) that the sets of \sol-defined
structures are unions of equivalence classes of isomorphism
(Def. \ref{def:isomorphism}).

In this paper we are concerned with the following definability problems:

\begin{table}[t!]
  \vspace*{-\baselineskip}
  \begin{center}
  \begin{tabular}{|c|c||c|c||c|c||c|c|}
    \hline
    $\subseteq$ & $\subseteq^k$ & \multicolumn{2}{|c||}{\slr} & \multicolumn{2}{|c||}{\mso} & \multicolumn{2}{|c|}{\sol} \\
    \hline
    \hline
    \multicolumn{2}{|c||}{\slr} & \checkmark & ? & $\times$ (\S\ref{sec:slr-mso}) & $\times$ (\S\ref{sec:slr-mso}) & \checkmark (\S\ref{sec:slr-so}) & \checkmark (\S\ref{sec:remaining}) \\
    \hline
    \multicolumn{2}{|c||}{\mso} & $\times$ (\S\ref{sec:mso-slr}) & \checkmark (\S\ref{sec:k-mso-slr}) & \checkmark & \checkmark (\S\ref{sec:remaining}) & \checkmark & \checkmark (\S\ref{sec:remaining}) \\
    \hline
    \multicolumn{2}{|c||}{\sol} & $\times$ (\S\ref{sec:mso-slr}) & ? & $\times$ (\S\ref{sec:remaining}) & $\times$ (\S\ref{sec:remaining}) & \checkmark & \checkmark (\S\ref{sec:remaining}) \\
    \hline
  \end{tabular}
  \end{center}
  \caption{A comparison of \slr, \mso\ and \sol\ in terms of
    expressiveness, where $\checkmark$ means that the inclusion holds,
    $\times$ means it does not and $?$ denotes an open problem.}
  \label{tab:expressiveness}
  \vspace*{-2\baselineskip}
\end{table}

\begin{definition}\label{def:definability}
  \begin{enumerate}
  \item[\slr\ $\subseteq$ \solmso:] Let $\phi$ and $\asid$ be a
    \slr\ sentence and SID, respectively. Does there exist a
    \solmso\ sentence $\psi$ such that the family of structures
    \slr-defined by $(\phi,\asid)$ is also \sol-defined by $\psi$?
  \item[\solmso\ $\subseteq$ \slr:] Let $\psi$ be a
    \solmso\ sentence. Does there exist a \slr\ sentence $\phi$ and a
    SID $\asid$, such that the family of structures \sol-defined by
    $\psi$ is \slr-defined by $(\phi,\asid)$?
  \end{enumerate}
\end{definition}
Additionally, we consider the \emph{treewidth-bounded} versions
of the above problems:

\begin{definition}\label{def:k-definability}
  \begin{enumerate}
  \item[\slr\ $\subseteq^k$ \solmso:] Let $\phi$ and $\asid$ be a
    \slr\ sentence and SID, respectively and $k\geq1$ be an
    integer. Does there exist a \solmso\ sentence $\psi$ such that the
    family of structures of treewidth at most $k$ \slr-defined by
    $(\phi,\asid)$ is \sol-defined by $\psi$?
  \item[\solmso\ $\subseteq^k$ \slr:] Let $\psi$ be a \solmso\ sentence and
    $k \geq 1$ be an integer. Does there exist a \slr\ sentence $\phi$
    and an SID $\asid$, such that the family of structures of treewidth
    at most $k$ that is \solmso-defined by $\psi$ is \slr-defined by
    $(\phi,\asid)$?
  \end{enumerate}
\end{definition}
For symmetry, we consider also the problems \mso\ $\subseteq^{(k)}$
\sol, \sol\ $\subseteq^{(k)}$ \mso. Table \ref{tab:expressiveness}
summarizes our results, with references to the sections in the paper
where the proofs (for the non-trivial ones) can be found, and the
remaining open problems.

\section{\slr\ $\not\subseteq^{(k)}$ \mso}
\label{sec:slr-mso}

We prove that \slr\ $\not\subseteq^{k}$ \mso, by exhibiting an
\slr-definable family of structures of treewidth $k=1$ that is not
\mso-definable. Note that this also implies that \slr\ $\not\subseteq$
\mso, in general.

The main idea is to encode words over a binary alphabet $\alphabet =
\set{a,b}$ by lists (connected acyclic graphs in which each vertex has
at most one incoming and at most one outgoing edge) whose vertices are
labeled by symbols from $\alphabet$. We encode a language $L$ as a
family of $\alphabet$-labeled lists, using the intuitive encoding of a
word $w = a_1 \ldots a_n$ by the structure $\struc_w$, over the
signature $\set{\vertset,\edgerel,\prel_a,\prel_b,\bcst,\ecst}$, such
that $\struc_w(\vertset) = \set{v_1,\ldots,v_n}$, $\struc_w(\edgerel)
= \set{(v_i,v_{i+1}) \mid i \in \interv{1}{n-1}}$,
$\struc_w(\prel_\alpha) = \set{v_i \mid i \in \interv{1}{n},~ a_i =
  \alpha}$, for all $\alpha \in \alphabet$, $\struc_w(\bcst)=v_1$ and
$\struc_w(\ecst)=v_n$. Note that $\twof{\struc_w} = 1$, for any word
$w \in \alphabet^*$.

A \emph{context-free grammar} $G = (N,\mathcal{X},\Delta)$ consists of
a finite set $N$ of \emph{nonterminals}, a start symbol $\mathcal{X}
\in N$ and a finite set $\Delta$ of \emph{productions} of the form
$\mathcal{Y} \rightarrow w$, where $\mathcal{Y} \in N$ and $w \in (N
\cup \alphabet)^*$. Given finite strings $u, v \in (N \cup \alphabet)^*$,
the relation $u \rhd v$ replaces a nonterminal $\mathcal{Y}$ of $u$ by
the right-hand side $w$ of a production $\mathcal{Y} \rightarrow w$
and $\rhd^*$ denotes the reflexive and transitive closure of
$\rhd$. The \emph{language} of $G$ is the set $\lang{G} \isdef \set{w
  \in \alphabet^* \mid \mathcal{X} \rhd^* w}$. A language $L$ is
\emph{context-free} if and only if there exists a context-free grammar
$G$, such that $L = \lang{G}$.

A language $L$ is \emph{recognizable} if and only if $L = h^{-1}(C)$,
where $C \subseteq \monoid$ and $h$ is a monoid morphism between the
monoid $(\alphabet^*,\cdot)$ of words with concatenation and a finite
monoid $(\monoid,\odot)$. It is known that a language is recognizable
if and only if it is definable by an \wss\ formula, i.e. an
\mso\ formula over the signature
$\set{\mathfrak{p}_a,\mathfrak{p}_b,\asucc}$, where $\mathfrak{p}_a$,
$\mathfrak{p}_b$ are unary predicates interpreted in every word $w$ of
length $n$ as the subsets of $\interv{1}{n}$ consisting of the
positions in the word labeled by $a$ and $b$, respectively and
$\asucc$ is a function symbol interpreted in every word as the
successor in $\nat$ (i.e., $\asucc(n)=n+1$) (see e.g.,
\cite{KhoussainovNerode} for a textbook presentation of this
result). It is also well-known that every recognizable language is
context-free but not viceversa e.g., $\set{a^n b^n \mid n \in \nat}$
is context-free but not recognizable.

\begin{propositionE}\label{prop:cfg-slr}
  Given a context-free grammar $G = (N,\mathcal{X},\Delta)$, there
  exists an SID $\asid_G$ and a binary predicate symbol
  $\apred_\mathcal{X} \in \defnof{\asid_G}$, such that $w \in \lang{G}
  \iff \struc_w \models_{\asid_G} \apred_\mathcal{X}(\bcst, \ecst)$,
  for all $w \in \alphabet^*$.
\end{propositionE}
\begin{proofE}
  Assume w.l.o.g that the context-free grammar $G$ does not produce
  the empty word and that it is in Greibach normal form i.e., contains
  only production rules of the form $\mathcal{Y}_0 \rightarrow \alpha
  \mathcal{Y}_1 \ldots \mathcal{Y}_i$, where $\mathcal{Y}_0, \ldots,
  \mathcal{Y}_i \in N$, for some $i \geq 0$ and some $\alpha \in
  \alphabet$. For each nonterminal $\mathcal{Y}$, we consider a binary
  relation symbol $\apred_\mathcal{Y}(x_1,x_2)$ and for each
  production rule as above, we consider a rule:
  \begin{align*}
    \apred_{\mathcal{Y}_0}(x_1,x_2) \leftarrow &
    ~\exists y_1 \ldots \exists y_{i} ~.~ \vertset(x_1) *
    \prel_\alpha(x_1) * \edgerel(x_1,y_1) *
    \apred_{\mathcal{Y}_1}(y_1,y_2) * \edgerel(y_2,y_3) * \ldots \\
    & \quad\quad * \edgerel(y_{i-1},y_i) * \apred_{\mathcal{Y}_i}(y_{i},x_2)
  \end{align*}
  if $i\geq1$ and \(\apred_{\mathcal{Y}_0}(x_1,x_2) \leftarrow
  \vertset(x_1) * \prel_\alpha(x_1) \text{, if $i = 0$}\). Let $\asid$ be the set of
  the rules above and let $\langof{\mathcal{Y}}{G} \isdef \set{w \in
    \alphabet^* \mid \mathcal{Y} \rhd^* w}$, for all $\mathcal{Y} \in
  N$. Let $w = a_1 \ldots a_n \in \alphabet^*$ be any word and
  $\struc_w$ be the structure over the signature
  $\set{\vertset,\edgerel,\prel_a,\prel_b,\bcst,\ecst}$, such that
  $\struc_w(\vertset) = \set{v_1,\ldots,v_n}$, $\struc_w(\edgerel) =
  \set{(v_j,v_{j+1}) \mid j \in \interv{1}{n-1}}$,
  $\struc_w(\prel_\alpha) = \set{v_j \mid j \in \interv{1}{n},~ a_j =
    \alpha}$, for all $\alpha \in \alphabet$, $\struc_w(\bcst)=v_1$
  and $\struc_w(\ecst)=v_n$. We prove that $\mathcal{Y} \rhd^* w \iff
  (\struc_w,\store) \models_\asid \apred_\mathcal{Y}(x,y)$, for any
  word $w \in \alphabet^*$ and any nonterminal $\mathcal{Y} \in N$,
  where $\store$ is any store such that $\store(x)=\struc_w(\bcst)$
  and $\store(y)=\struc_w(\ecst)$.

  \noindent ``$\Rightarrow$'' By induction on the length $n\geq1$ of
  the derivation of $w$ from $\mathcal{Y}$. The base case $n=1$
  corresponds to a production rule $\mathcal{Y} \rightarrow \alpha$ in
  $G$ that yields the rule $\apred_{\mathcal{Y}}(x_1,x_2) \leftarrow
  \vertset(x_1) * \prel_\alpha(x_1)$ in $\asid$ and $\struc_\alpha
  \models_\asid \apred_{\mathcal{Y}}(\bcst,\ecst)$ follows. For the
  inductive step $n>1$, we assume w.l.o.g. that the derivation is
  ordered such that each nonterminal is fully expanded before another
  nonterminal from the same rule (every derivation of a context-free
  grammar can be reordered in this way). Let $\mathcal{Y}_0
  \rightarrow \alpha \mathcal{Y}_1 \ldots \mathcal{Y}_i$ be the first
  rule of the derivation and let $\mathcal{Y}_j \rhd^* w_j$ be the
  sub-derivations of $\mathcal{Y}_1, \ldots, \mathcal{Y}_i$,
  respectively.
  Then $w = \alpha w_1 \ldots w_i$.
  We can now choose structures $\struc_j$, with $0 \le j \le i$, such that
  \begin{itemize}
  \item $\struc_w = \struc_0 \comp \struc_1 \comp \cdots \comp \struc_i$,
  \item $\struc_0$ is a structure with  $\struc_0(\vertset) = \set{v_0}$,   $\struc_0(\prel_\alpha) = \set{v_0}$,   $\struc_0(\edgerel)=\set{(v_j,v_{j+1}) \mid j \in \interv{0}{n-1}}$, $\struc_0(\arel)=\emptyset$, for all $\arel \not\in \set{\vertset,\prel_\alpha,\edgerel}$, $\struc_0(\bcst)=v_0$ and $\struc_0(\ecst)=v_n$ for some set of pairwise different nodes $\{v_0,\ldots,v_n\}$.
  \end{itemize}
  We denote by $\struc_j'$, for $1 \le j \le i$, the structure that is identical to $\struc_j$ except that we set $\struc_j'(\bcst) = v_j$ and $\struc_j'(\ecst) = v_{j+1}$, where we set $v_{n+1} = \struc_w(\ecst)$.
  We recognize that $\struc_j'$ is isomorphic to $\struc_{w_j}$.
  By the inductive hypothesis, we have $\struc_{w_j}
  \models_\asid \apred_{\mathcal{Y}_j}(\bcst,\ecst)$ for all $1 \le j \le i$.
  By Prop. \ref{prop:sl-iso} we get that  $\struc_j'
  \models_\asid \apred_{\mathcal{Y}_j}(\bcst,\ecst)$.
  Let $\store$ be a store with $\store(y_j) = v_j$ for all $1 \le j \le i$.
  We now recognize that
  \[(\struc_0 \comp \struc_1 \comp \ldots \comp \struc_i, \store) \models_\asid
  \vertset(\bcst) * \prel_\alpha(\bcst) * \edgerel(\bcst,y_1) *
  \apred_{\mathcal{Y}_1}(y_1,y_2) * \edgerel(y_2,y_3) * \ldots *
  \edgerel(y_{i-1},y_i) * \apred_{\mathcal{Y}_i}(y_i,\ecst).\]
  Hence, $\struc_w \models_\asid
  \apred_{\mathcal{Y}}(\bcst,\ecst)$.

  \noindent ``$\Leftarrow$'' By induction on the definition of the
  satisfaction relation. In the base case, we have
  $\struc_\alpha \models \vertset(\bcst) * \prel_\alpha(\bcst)$, leading to
  $\mathcal{Y} \rhd^* \alpha$, for a rule $\mathcal{Y} \rightarrow
  \alpha$ of $G$.
  For the inductive step, assume that
  \[(\struc_w,\store) \models_\asid \vertset(\bcst) * \prel_\alpha(\bcst) * \edgerel(\bcst,y_1)
  * \apred_{\mathcal{Y}_1}(y_1,y_2) * \edgerel(y_2,y_3) * \ldots *
  \edgerel(y_{i-1},y_i) * \apred_{\mathcal{Y}_i}(y_i,\ecst)\] for some
  store $\store$ with $\dom{\store} = \{y_1,\ldots,y_i\}$, where
  $\apred_{\mathcal{Y}_0}(x_1,x_2) \leftarrow \exists y_1 \ldots
  \exists y_{i} ~.~ \vertset(x_1) * \prel_\alpha(x_1) *
  \edgerel(x_1,y_1) * \apred_{\mathcal{Y}_1}(y_1,y_2) *
  \edgerel(y_2,y_3) * \ldots * \edgerel(y_{i-1},y_i) *
  \apred_{\mathcal{Y}_i}(y_i,x_2)$ is a rule of $\asid$, for some $i
  \geq 1$.  Then $G$ has a rule $\mathcal{Y} \rightarrow \alpha
  \mathcal{Y}_1 \ldots \mathcal{Y}_i$.  Moreover, there exist
  structures $\struc_j$ with $\struc_0 \comp \struc_1 \comp \ldots
  \comp \struc_i = \struc_w$, $(\struc_0,\store) \models
  \vertset(\bcst) * \prel_\alpha(\bcst) * \edgerel(\bcst,y_1) *
  \edgerel(y_2,y_3) * \ldots * \edgerel(y_{i-1},y_i)$,
  $(\struc_j,\store) \models_\asid
  \apred_{\mathcal{Y}_j}(y_j,y_{j+1})$ for all $j \in \interv{1}{i-1}$
  and $(\struc_i,\store) \models_\asid
  \apred_{\mathcal{Y}_i}(y_i,\ecst)$.  Let $\struc'_j$ be the
  structures that agree with $\struc_j$, except that $\struc'_j(\bcst)
  = \store'(y_j)$, $\struc'_j(\ecst)=\store(y_{j+1})$, for all $j \in
  \interv{1}{i-1}$, and $\struc'_i(\ecst)=\struc_w(\ecst)$.  It is
  easy to show that there exist words $w_1, \ldots, w_i$, such that $w
  = \alpha w_1 \ldots w_i$ and $\struc_{w_j} \iso \struc'_j$, for all
  $j \in \interv{1}{i}$.  By Prop. \ref{prop:sl-iso}, we obtain
  $\struc_{w_j} \models_\asid \apred_{\mathcal{Y}_j}(\bcst,\ecst)$,
  hence $\mathcal{Y}_j \rhd^* w_j$, by the inductive hypothesis, thus
  leading to $\mathcal{Y} \rhd^* w$.
\end{proofE}
Let $L$ be a non-recognizable context-free language and suppose that
there exists an \mso\ formula $\phi_L$, over the signature
$\set{\vertset,\edgerel,\prel_a,\prel_b,\bcst,\ecst}$, that defines
the family $\mathcal{S}_L \isdef \set{\struc_w \mid w \in L}$. Then
there exists also a \wss\ formula $\psi_L$ that defines $L$. Note that
$\psi_L$ can be obtained directly from $\phi_L$ by replacing each atom
$\edgerel(\xi,\chi)$ by $\asucc(\xi)=\chi$ and each atom
$\prel_\alpha(\xi)$ by $\mathfrak{p}_\alpha(\xi)$. But then the
language $L$ is recognizable, which contradicts with the choice of
$L$. Moreover, we have $\set{\twof{\struc} \mid \struc \in
  \mathcal{S}_L} = \set{1}$, proving that \slr\ $\not\subseteq^k$
\mso, for any given $k \geq 1$.

\section{\mso\ $\not\subseteq$ \slr}
\label{sec:mso-slr}

We prove that the \mso-definable indexed family of cliques
$\set{\clique{n} \mid n \in \nat}$ is not \slr-definable. Because
\mso\ is a fragment of \sol, this also implies \sol\ $\not\subseteq$
\slr. First, observe that $\set{\clique{n} \mid n \in \nat}$ is
defined by the \mso\ formula $\forall x \forall y ~.~ \vertset(x)
\wedge \vertset(y) \wedge x \neq y \rightarrow \edgerel(x,y)$.
\ifLongVersion The definition of square grids in \mso\ is given in
\cite[\S5.2.3]{courcelle_engelfriet_2012}.  \fi As shown in
Proposition \ref{prop:strictly-unbounded}, this family is strictly
treewidth-unbounded. We prove that \slr\ cannot define strictly
treewidth-unbounded families of structures, by showing the existence
of an integer $k\geq1$, depending only on $\phi$ and $\asid$, such
that for each model of $\phi$ there exists another model of $\phi$ of
greater or equal size and treewidth bounded by $k$ (Proposition
\ref{prop:slr-not-strictly-unbounded}). This is a consequence of the
fact that each SID can be transformed into an equivalent SID in which
the variables that occur existentially quantified in the rules of
$\asid$ are not constrained by equalities.

\begin{definition}\label{def:normalized-sid}
  A rule $\apred(x_1, \ldots, x_{\arityof{\apred}}) \leftarrow \exists
  y_1 \ldots \exists y_n ~.~ \psi$, where $\psi$ is a quantifier-free
  formula, is \emph{normalized} if and only if no equality atom $x =
  y$ occurs in $\psi$, for distinct variables $x,y \in \set{x_1,
    \ldots, x_{\arityof{\apred}}} \cup \set{y_1, \ldots, y_n}$. An SID
  is \emph{normalized} if and only if it contains only normalized
  rules.
\end{definition}

\begin{lemmaE}\label{lemma:normalized-sid}
  Given an SID $\asid$, one can build a normalized SID $\asid'$ such
  that $\defnof{\asid} \subseteq \defnof{\asid'}$ and for each
  structure $\struc$ and each predicate atom $\apred(\xi_1, \ldots,
  \xi_{\arityof{\apred}})$, we have $\struc \models_\asid \exists
  \xi_{i_1} \ldots \exists \xi_{i_n} ~.~ \apred(\xi_1, \ldots,
  \xi_{\arityof{\apred}}) \iff \struc \models_{\asid'} \exists
  \xi_{i_1} \ldots \exists \xi_{i_n} ~.~ \apred(\xi_1, \ldots,
  \xi_{\arityof{\apred}})$, where $\set{\xi_{i_1}, \ldots, \xi_{i_n}}
  = \set{\xi_1, \ldots, \xi_{\arityof{\apred}}} \cap \vars$.
\end{lemmaE}
\begin{proofE}
  Let $\asid$ be an SID. For each predicate $\apred \in \defnof{\asid}$
  and each partition $\set{I_1, \ldots, I_k}$ of
  $\interv{1}{\arityof{\apred}}$, we consider a fresh predicate
  $\apred_{I_1,\ldots,I_k}$ of arity $k\geq1$, not in
  $\defnof{\asid}$. Let $\asid'$ be the SID obtained from $\asid$ by
  introducing, for each rule:
  \begin{align}
    \apred(x_1, \ldots, x_{\arityof{\apred}}) \leftarrow \exists y_1 \ldots \exists y_m ~.~
    \phi * \Asterisk_{\ell=1}^h \bpred^\ell(z_{\ell,1}, \ldots,
    z_{\ell,\arityof{\bpred^\ell}}) \in \asid \label{rule:stem}
  \end{align} where $\phi$ is a
  quantifier- and predicate-free formula and for each equivalence
  relation $\approx$ on the set of variables $\set{x_1, \ldots,
    x_{\arityof{\apred}}} \cup \set{y_1,\ldots,y_m}$ that is
  compatible with all equalities in $\phi$ i.e., $x = y$ occurs in $\phi$ only if $x \approx y$,
  the following rules:
  \begin{align}
    \apred_{I_1,\ldots,I_k}(x_1, \ldots, x_k) \leftarrow & ~\big(\exists y_{j_1} \ldots \exists y_{j_n}
    ~.~ \psi * \Asterisk_{\ell=1}^k \bpred^\ell_{J^\ell_1,\ldots,J^\ell_{s^\ell}}(z_{r_{\ell,1}}, \ldots,
    z_{r_{\ell,s^\ell}})\big) \label{rule:one} \\[-2mm]
    & ~~[x_{i_1}/x_1, \ldots, x_{i_k}/x_k] \nonumber \\[2mm]
    \apred(x_1, \ldots, x_{\arityof{\apred}}) \leftarrow &
    ~\apred_{I_1,\ldots,I_k}(x_{i_1}, \ldots, x_{i_k}) \label{rule:two}
  \end{align}
  where:
  \begin{compactitem}
  \item $\approx$ induces the partitions $\set{I_1, \ldots, I_k}$ of
    $\interv{1}{\arityof{\apred}}$ and $\set{J^\ell_1, \ldots,
    J^\ell_{s^\ell}}$ of $\interv{1}{\arityof{\bpred^\ell}}$, for each
    $\ell \in \interv{1}{k}$,
  \item $x_{i_j}$ and $z_{r_{\ell,s^\ell}}$ are the first in their
    $\approx$-equivalence classes, respectively, in the total order
    $x_1 < \ldots < x_{\arityof{\apred}} < y_1 <
    \ldots < y_m$,
  \item $\psi$ is obtained from $\phi$ by replacing each variable $x$,
    such that $x \approx x_{i_j}$ with $x_{i_j}$, respectively each
    $z$, such that $z \approx z_{r_{\ell,s^j}}$ with
    $z_{r_{\ell,s^j}}$, and removing the trivial equalities of the
    form $x = x$,
  \item the quantifier prefix $\exists y_{j_1} \ldots \exists y_{j_n}$
    is the result of eliminating from $\exists y_1 \ldots \exists y_m$
    the variables that do not occur in $\fv{\psi} \cup
    \bigcup_{\ell=1}^h \set{z_{\ell,r_1}, \ldots,
      z_{r_{\ell,s^\ell}}}$.
  \end{compactitem}
  In particular, one can remove from $\asid'$ the rules containing
  unsatisfiable disequalities of the form $x \neq x$, obtained from
  the above transformations. We are left with proving the equivalence
  from the statement.

  \noindent
  ``$\Rightarrow$'' Assume that $(\struc,\store) \models_\asid
  \apred(\xi_1, \ldots, \xi_{\arityof{\apred}})$ for a store $\store$,
  and let $\approx$ be the equivalence relation over $x_1 < \ldots <
  x_{\arityof{\apred}}$ defined as $\xi_i \approx \xi_j \iff
  (\struc,\store)(\xi_i) = (\struc,\store) (\xi_j)$.
  We now prove by induction that $(\struc,\store) \models_\asid
  \apred(\xi_1, \ldots, \xi_{\arityof{\apred}})$ implies
  $(\struc,\store) \models_{\asid'} \apred_{I_1, \ldots,
    I_k}(\xi_{i_1}, \ldots, \xi_{i_k})$, where $\set{I_1, \ldots,
    I_k}$ are the partitions of $\interv{1}{\arityof{\apred}}$ induced
  by $\approx$ and the $\xi_{i_j}$ are minimal representatives of
  $I_j$.  Since $(\struc,\store) \models_\asid \apred(\xi_1, \ldots,
  \xi_{\arityof{\apred}})$, there is a rule (\ref{rule:stem}) in
  $\asid$, a store $\store'$, that agrees with $\store$ over $\xi_{1},
  \ldots, \xi_{\arityof{\apred}}$ and structures $\struc_0 \comp
  \ldots \comp \struc_h = \struc$, such that $(\struc_0,\store')
  \models \phi\overline{s}$ and $(\struc_i,\store') \models_\asid
  \bpred^\ell(z_{\ell,1},\ldots,z_{\ell,\arityof{\bpred^\ell}})\overline{s}$,
  for all $\ell \in \interv{1}{h}$, where $\overline{s} \isdef
  [x_{1}/\xi_{1}, \ldots, x_{n}/\xi_{n}]$ is the substitution that
  replaces the formal parameters by terms. Let $\approx'$ be the
  equivalence over $x_1 < \ldots < x_{\arityof{\apred}} < y_1 < \ldots
  < y_m$ defined as $x \approx y \iff \store'(x) = \store'(y)$. By the
  inductive hypothesis, we have $(\struc_i,\store') \models_{\asid'}
  \bpred^\ell_{J_1, \ldots,
    r_{s^\ell}}(z_{J_{\ell,1}},\ldots,z_{r_{\ell,s^\ell}})\overline{s}$,
  where $z_{r_{\ell,1}},\ldots,z_{r_{\ell,s^\ell}}$ is the sequence of
  minimal representatives wrt $\approx'$.  Since $\approx' \ \supseteq
  \ \approx$, there exists a rule of type (\ref{rule:one}) in $\asid'$
  allowing to infer that $(\struc,\store) \models_{\asid'}
  \apred_{I_1, \ldots, I_k}(\xi_{i_1}, \ldots, \xi_{i_k})$.
  We finally obtain $(\struc,\store) \models_{\asid'} \apred(\xi_1, \ldots,
  \xi_{\arityof{\apred}})$, by a rule of type (\ref{rule:two}).

  \noindent''$\Leftarrow$'' Assume that $(\struc,\store)
  \models_{\asid'} \apred(\xi_1, \ldots, \xi_{\arityof{\apred}})$.
  Then, by a rule of type (\ref{rule:two}) from $\asid'$, we must have
  $(\struc,\store) \models_{\asid'} \apred_{I_1, \ldots,
    I_k}(\xi_{i_1}, \ldots, \xi_{i_k})$.  We now prove by induction
  that $(\struc,\store) \models_{\asid'} \apred_{I_1, \ldots,
    I_k}(\xi_{i_1}, \ldots, \xi_{i_k})$ implies that $(\struc,\store')
  \models_{\asid} \apred(\xi_1, \ldots, \xi_{\arityof{\apred}})$,
  where $\store'$ is the store that maps each $\xi_j \in \vars$, such
  that $j \in I_j$, into $(\struc,\store)(\xi_{i_j})$.  Assume that
  $(\struc,\store'') \models_{\asid'} (\psi * \Asterisk_{\ell=1}^h
  \bpred^\ell_{J^\ell_1,\ldots,J^\ell_{s^\ell}}(z_{r_{\ell,1}},
  \ldots, z_{r_{\ell,s^\ell}}))\overline{s}$, by a rule of type
  (\ref{rule:one}), where $\store''$ is a store that agrees with
  $\store$ over $\xi_{i_1}, \ldots, \xi_{i_k}$ and $\overline{s}
  \isdef [x_1/\xi_{i_1}, \ldots, x_k/\xi_{i_k}]$ is a
  substitution. Then there exists structures $\struc_0 \comp \ldots
  \comp \struc_h = \struc$, such that $(\struc_0,\store'') \models
  \psi\overline{s}$ and $(\struc_\ell,\store'') \models_{\asid'}
  \bpred^\ell_{J^\ell_1,\ldots,J^\ell_{s^\ell}}(z_{r_{\ell,1}},
  \ldots, z_{r_{\ell,s^\ell}})\overline{s}$, for all $\ell \in
  \interv{1}{h}$. By the definition of $\asid'$, there exists a rule
  of type (\ref{rule:stem}) in $\asid$ and a corresponding equivalence
  relation $\approx$ over $x_1 < \ldots < x_{\arityof{\apred}} < y_1 <
  \ldots < y_m$, which induces the partitions $\set{I_1, \ldots, I_k}$
  of $\interv{1}{\arityof{\apred}}$ and $\set{J^\ell_1, \ldots,
    J^\ell_{s^\ell}}$ of $\interv{1}{\arityof{\bpred^\ell}}$, for each
  $\ell \in \interv{1}{k}$, and $x_{i_j}$ and $z_{r_{\ell,s^\ell}}$
  are the first in their $\approx$-equivalence classes.  We can now
  choose a store $\store'''$, such that:
  \begin{compactitem}
  \item $(\struc,\store''')(\xi_j) =  (\struc,\store'')(\xi_{i_j})$, for all $j \in \interv{1}{k}$, and
  \item $(\struc,\store''')(z_j\overline{s}) = (\struc,\store'')(z_{r_{\ell,q}}\overline{s})$ if $j \in J^\ell_q$, for all $\ell\in\interv{1}{h}$ and $q \in    \interv{1}{s^\ell}$.
  \end{compactitem}
  By the definition of $\psi$, we have $(\struc_0,\store''') \models
  \phi$. By the inductive hypothesis, we obtain
  $(\struc_\ell,\store''') \models_\asid \bpred^\ell(z_{\ell,1},
  \ldots, z_{\ell,\arityof{\bpred^\ell}})$, for all
  $\ell\in\interv{1}{h}$.
  Hence $(\struc,\store''') \models_\asid
  \apred(\xi_1,\ldots,\xi_{\arityof{\apred}})$, by a rule of type
  (\ref{rule:stem}).
\end{proofE}
A consequence of the above result is that, in the absence of equality
constraints in a normalized SID, each existentially quantified
variable instantiated by the inductive definition of the satisfaction
relation can be assigned a distinct element of the universe. We write
$(\struc,\store) \imodels{U}{\asid} \phi$ if and only if the
satisfaction relation $(\struc,\store) \models_\asid \phi$ can be
established by considering stores that are injective over the
interpretation of existentially quantified variables and map these
variables into elements from an infinite set $U \subseteq
\universe$. More precisely, the inductive definition of
$\imodels{U}{\asid}$ is the same as the definition of $\models_\asid$,
except for the following cases:
\vspace*{-.5\baselineskip}
\[\begin{array}{rclcl}
(\struc,\store) & \imodels{U}{\asid} & \phi_1 * \phi_2 & \iff &
\text{there exist structures } \struc_1 \comp \struc_2 = \struc \text{ and infinite sets} \\
&&&&U_1 \cap U_2 = \emptyset,~ U_1 \cup U_2 \subseteq U \text{, such that }
(\struc_i,\store) \imodels{U_i}{\asid} \phi_i, \\
&&&& \text{for all } i = 1,2 \\
(\struc,\store) & \imodels{U}{\asid} & \exists x ~.~ \phi & \iff &
(\struc,\store[x\leftarrow u]) \models_\asid \phi \text{, for some } u
\in U \setminus \store(\fv{\phi}\setminus\set{x})
\end{array}\]

\vspace*{-.5\baselineskip}
\begin{lemmaE}\label{lemma:injective-model}
  Given a normalized SID $\asid$, a predicate atom
  $\apred(\xi_1,\ldots,\xi_{\arityof{\apred}})$ and an infinite set $U
  \subseteq \universe$, for each structure-store pair
  $(\struc,\store)$, such that $(\struc,\store) \models_\asid
  \apred(\xi_1,\ldots,\xi_{\arityof{\apred}})$, there exists a
  structure $\struc'$, such that $(\struc',\store) \imodels{U}{\asid}
  \apred(\xi_1,\ldots,\xi_{\arityof{\apred}})$ and
  $\cardof{\Dom{\struc}} \leq \cardof{\Dom{\struc'}}$.
\end{lemmaE}
\begin{proofE} The structure $\struc'$ is built inductively on the definition
  of the satisfaction relation $(\struc,\store) \models_\asid
  \apred(x_1,\ldots,x_{\arityof{\apred}})$. Since no existentially
  quantified variable is constrained by equality during this
  derivation, one can use the definition of $\imodels{U}{\asid}$
  instead, thus ensuring that $(\struc',\store) \imodels{U}{\asid}
  \apred(x_1,\ldots,x_{\arityof{\apred}})$. Moreover, since the values
  of all existentially quantified variables are pairwise distinct in
  $\struc'$, it follows that $\cardof{\Dom{\struc}} \leq
  \cardof{\Dom{\struc'}}$.
\end{proofE}
We show that the models defined on injective stores have bounded treewidth:

\begin{lemmaE}\label{lemma:injective-bounded}
  Given a normalized SID $\asid$ and a predicate atom $\apred(\xi_1,
  \ldots, \xi_{\arityof{\apred}})$, we have $\twof{\struc} \leq k$,
  for each structure-store pair $(\struc,\store)$ such that
  $(\struc,\store) \imodels{U}{\asid}
  \apred(\xi_1,\ldots,\xi_{\arityof{\apred}})$, for some infinite set
  $U \subseteq \universe$, where $k\geq1$ is a constant depending only
  on $\asid$.
\end{lemmaE}
\begin{proofE}
  Let $k$ be the maximal number of variables that occurs free or bound
  in the right-hand side of a rule from $\asid$. Given an infinite set
  $U \subseteq \universe$, we build a tree decomposition $\tree =
  (\tnodes,\tedges,r,\alabel)$ of $\struc$, inductively on the
  definition of the satisfaction relation $(\struc,\store)
  \imodels{U}{\asid} \apred(\xi_1,\ldots,\xi_{\arityof{\apred}})$,
  such that $\bigcup_{n \in \tnodes} \alabel(n) \subseteq U \cup
  \set{(\struc,\store)(\xi_1), \ldots,
    (\struc,\store)(\xi_{\arityof{\apred}})}$. Assume that
  $(\struc,\store) \imodels{U}{\asid}
  \apred(\xi_1,\ldots,\xi_{\arityof{\apred}})$ is the consequence of a
  rule $\apred(x_1,\ldots,x_{\arityof{\apred}}) \leftarrow \exists y_1
  \ldots \exists y_m ~.~ \psi * \Asterisk_{\ell=1}^k
  \bpred_\ell(z^\ell_1, \ldots, z^\ell_{\arityof{\bpred_\ell}})$ from
  $\asid$, where $\psi$ is a quantifier- and predicate-free formula,
  such that:
  \[(\struc,\store') \imodels{U}{\asid} \exists y_1
  \ldots \exists y_m ~.~ \psi * \Asterisk_{\ell=1}^k
  \bpred_\ell(z^\ell_1, \ldots, z^\ell_{\arityof{\bpred_\ell}})\]
  where $\store'$ is the store that maps $x_i$ into $\store(\xi_i)$,
  for all $i \in \interv{1}{\arityof{\apred}}$. Then there exists a
  store $\store''$ that agrees with $\store'$ over $x_1, \ldots,
  x_{\arityof{\apred}}$, such that $\store''(y_1), \ldots,
  \store''(y_m) \in U$ are pairwise distinct, and structures $\struc_0
  \comp \struc_1 \comp \ldots \comp \struc_k = \struc$, such that
  $(\struc_0,\store'') \models \psi$ and $(\struc_\ell,\store'')
  \imodels{U_\ell}{\asid} \bpred_\ell(z^\ell_1, \ldots,
  z^\ell_{\arityof{\bpred_\ell}})$, for all $\ell \in \interv{1}{k}$,
  where $U_1, \ldots, U_k$ is a partition of $U$ into infinite
  sets. Note that, because $U$ is infinite, such a partition always
  exists. By the inductive hypothesis, there exists a tree
  decomposition $\tree_\ell =
  (\tnodes_\ell,\tedges_\ell,r_\ell,\alabel_\ell)$ of $\struc_\ell$,
  such that $\twof{\tree_\ell} \leq k$ and $\bigcup_{n \in
    \tnodes_\ell} \alabel_\ell(n) \subseteq U_\ell \cup
  \set{\store''(z^\ell_1), \ldots,
    \store''(z^\ell_{\arityof{\bpred_\ell}})}$, for each $\ell \in
  \interv{1}{k}$.

  We define the tree decomposition $\tree =
  (\tnodes,\tedges,r,\alabel)$ such that the subtrees rooted in the
  children of the root are $\tree_1, \ldots, \tree_k$ and
  $\alabel(r)=\set{\store''(x_1), \ldots,
    \store''(x_{\arityof{\apred}})} \cup \set{\store''(y_1), \ldots,
    \store''(y_m)}$. Then, for each relation atom
  $\arel(z_1,\ldots,z_{\arityof{\arel}})$ that occurs in $\psi$, the
  set $\set{\store''(z_1), \ldots, \store''(z_{\arityof{\arel}})}$ is
  a subset of the label of the root, thus fulfilling point
  (\ref{it1:treewidth}) of Def. \ref{def:treewidth}. To check point
  (\ref{it2:treewidth}) of Def. \ref{def:treewidth}, let $u \in
  \alabel_i(n_i) \cap \alabel_j(n_j)$, where $n_i \in \tnodes_i$ and
  $n_j \in \tnodes_j$, for some $1 \leq i < j \leq k$. Since
  $(\struc_i,\store'') \imodels{U_i}{\asid}
  \bpred_i(z^i_1,\ldots,z^i_{\arityof{\bpred_i}})$,
  $(\struc_j,\store'') \imodels{U_j}{\asid}
  \bpred_j(z^j_1,\ldots,z^j_{\arityof{\bpred_j}})$ and $U_i \cap U_j =
  \emptyset$, we obtain that $u$ is not the image of an existentially
  quantified variable via $\store''$, hence $u = \store''(z)$, for
  some $z \in \set{z^i_1,\ldots,z^i_{\arityof{\bpred_i}}} \cap
  \set{z^j_1,\ldots,z^j_{\arityof{\bpred_j}}} \subseteq \set{\xi_1,
    \ldots, \xi_{\arityof{\apred}}} \cup \set{y_1, \ldots, y_m}$. Then
  $u \in \alabel(r)$, thus fulfilling point (\ref{it2:treewidth}) of
  Def. \ref{def:treewidth}.
\end{proofE}

The proof of \mso\ $\not\subseteq$ \slr\ relies on the following
result:

\begin{propositionE}\label{prop:slr-not-strictly-unbounded}
  Given an SID $\asid$ and a \slr\ sentence $\phi$, the family
  $\set{\struc \mid \struc \models_\asid \phi}$ is either finite or it
  has an infinite subfamily of bounded treewidth.
\end{propositionE}
\begin{proofE}
Given a sentence $\phi$, we introduce a fresh predicate $\apred_\phi$
of arity zero and consider the rule $\apred_\phi() \leftarrow
\phi$. Then, for each structure $\struc$, we have $\struc
\models_\asid \phi$ if and only if $\struc \models_{\asid_\phi}
\apred_\phi()$, where $\asid_\phi \isdef \asid \cup \set{\apred_\phi()
  \leftarrow \phi}$. Let $\asid'_\phi$ be the normalized SID, such
that $\struc \models_{\asid'_\phi} \apred_\phi()$, by Lemma
\ref{lemma:normalized-sid}. Given an infinite subset $U$ of
$\universe$, there exists a structure $\struc'$, such that $\struc'
\models^U_{\asid'_\phi} \apred_\phi()$ and $\cardof{\Dom{\struc}} \leq
\cardof{\Dom{\struc'}}$, by Lemma \ref{lemma:injective-model}. By
Lemma \ref{lemma:injective-bounded}, we also obtain $\twof{\struc'}
\leq k$, where $k$ depends only of $\asid'_\phi$ and hence of $\asid$
and $\phi$. If $\apred_\phi()$ has finitely many models, there is
nothing to prove. Otherwise, consider an infinite sequence $\struc_1,
\struc_2, \ldots$ of models of $\apred_\phi()$, such that
$\cardof{\Dom{\struc_i}} < \cardof{\Dom{\struc_{i+1}}}$, for all $i
\geq 1$. Then there exists a sequence of models $\struc'_1, \struc'_2,
\ldots$ of $\apred_\phi()$, such that $\cardof{\Dom{\struc_i}} \leq
\cardof{\Dom{\struc'_i}}$ and $\twof{\struc'_i} \leq k$, for all $i
\geq 1$. By going to a subsequence, if necessary, one can retrieve an
infinite treewidth-bounded family of models of $\phi$.
\end{proofE}

\section{\slr\ $\subseteq$ \sol}
\label{sec:slr-so}

Since \slr\ and \mso\ have incomparable expressivities (at least on
unbounded sets of structures), it is natural to ask for a logic that
subsumes them both. In this section, we prove that \sol\ is such a
logic. Since \mso\ is a syntactic subset of \sol, we have
\mso\ $\subseteq$\ \sol\ trivially.

In the rest of this section, we show that \slr\ $\subseteq$ \sol,
using the fact that each model of a predicate atom in \slr\ is built
according to an \emph{unfolding tree} indicating the partial order in
which the rules of the SID are used in the inductive definition of the
satisfaction relation\footnote{Unfolding trees are for SIDs what
  derivation trees are for context-free grammars.}. More precisely,
the model can be decomposed into pairwise disjoint substructures, each
being the model of the quantifier- and predicate-free subformula of a
rule in the SID, such that there is a one-to-one mapping between the
nodes of the tree and the decomposition of the model. We use
second-order variables to define the unfolding tree and the mapping
between the nodes of the unfolding tree and the tuples in the
interpretation of the relation symbols from the model. Finally, these
second-order variables are existentially quantified and the resulting
\sol\ formula describes the model alone, without the unfolding tree
that witnesses its decomposition according to the rules of the SID.

In the following, let $\signature \isdef \set{\arel_1, \ldots,
  \arel_\nrel, \acst_1, \ldots, \acst_\ncst}$ be the signature of
\slr\ and let $\asid \isdef \set{\arule_1, \ldots, \arule_\nrule}$ be
a given SID. Without loss of generality, for each relation symbol
$\arel_i \in \signature$, we assume that there is at most one
occurrence of a relation atom $\arel_i(y_1, \ldots,
y_{\arityof{\arel_i}})$ in each rule from $\asid$. If this is not the
case, we split the rule by introducing a new predicate symbol for each
relation atom with relation symbol $\arel_i$, until the condition is
satisfied.

We define unfolding trees formally. For a tree
$\tree=(\tnodes,\tedges,r,\alabel)$ and a vertex $n \in \tnodes$, we
denote by $\proj{\tree}{n}$ the subtree of $\tree$ whose root is $n$.
For a quantifier- and predicate-free formula $\phi$, we denote by
$\phi^n$ the formula in which every relation atom $\arel(x_1, \ldots,
x_{\arityof{\arel}})$ is annotated as $\arel^n(x_1, \ldots,
x_{\arityof{\arel}})$. Atoms $\arel^n(x_1, \ldots,
x_{\arityof{\arel}})$ (and consequently formulas $\phi^n$) have the
same semantics as atoms $\arel(x_1, \ldots, x_{\arityof{\arel}})$
(resp. formulas $\phi$); the annotations $\cdot^n$ purely serve as
explanatory devices in our construction (see the proof of
Proposition~\ref{prop:slr-so}) by tracking the node of the unfolding
tree where a given relation atom is introduced.

\begin{definition}\label{def:unfolding-tree}
  An \emph{unfolding tree} for a predicate atom $\apred(\xi_1, \ldots,
  \xi_{\arityof{\apred}})$ is a tree $\tree =
  (\tnodes,\tedges,r,\alabel)$ with labeling mapping $\alabel :
  \tnodes \rightarrow \asid$, such that $\alabel(r) \in
  \defn{\asid}{\apred}$ and, for each vertex $n \in \tnodes$, if
  $\bpred_1(z_{1,1}, \ldots, z_{1,\arityof{\bpred_1}}), \ldots,
  \bpred_h(z_{h,1}, \ldots, z_{h,\arityof{\bpred_h}})$ are the
  predicate atoms that occur in $\alabel(n)$, then $p_1, \ldots, p_h$
  are the children of $n$ in $\tree$, such that $\alabel(p_\ell) \in
  \defn{\asid}{\bpred_\ell}$, for all $\ell \in \interv{1}{h}$.

  An unfolding tree $\tree =
  (\tnodes,\tedges,r,\alabel)$ for a predicate atom $\apred(\xi_1, \ldots,
  \xi_{\arityof{\apred}})$ gives rise to a predicate-free formula defined inductively as follows:
  \vspace*{-.5\baselineskip}
  \begin{multline*}
    \charform{n}{\tree}{\apred(\xi_1, \ldots, \xi_{\arityof{\apred}})} \isdef  \\
  \left(\exists y_1 \ldots \exists y_m ~.~ \psi^r *
  \Asterisk_{\ell=1}^h
  \charform{p_\ell}{\proj{\tree}{p_\ell}}{\bpred_\ell(z_{\ell,1},
    \ldots, z_{\ell,\arityof{\bpred_\ell}})}\right) [x_1/\xi_1,
    \ldots, x_{\arityof{\apred}}/\xi_{\arityof{\apred}}],
  \end{multline*}

  \vspace*{-.5\baselineskip}\noindent
  where $\alabel(r)=\apred(x_1, \ldots, x_{\arityof{\apred}}) \leftarrow
  \exists y_1 \ldots \exists y_m ~.~ \psi * \Asterisk_{\ell=1}^h
  \bpred_\ell(z_{\ell,1}, \ldots, z_{\ell,\arityof{\bpred_\ell}})$,
  for a quantifier- and predicate-free formula $\psi$, $p_1, \ldots, p_h$ are the children of the root $r$ in $\tree$, corresponding to
  the predicate atoms $\bpred_1(z_{1,1}, \ldots,
  z_{1,\arityof{\bpred_1}}), \ldots, \bpred_h(z_{h,1}, \ldots,
  z_{h,\arityof{\bpred_h}})$, respectively.
\end{definition}
The unfolding trees of a predicate atom describe the set of models of
that predicate atom. The following lemma is standard and we include it
for self-containment reasons:

\begin{lemmaE}\label{lemma:unfolding-tree}
  For any structure-store pair $(\struc,\store)$, we have
  $(\struc,\store) \models_\asid \apred(\xi_1, \ldots,
  \xi_{\arityof{\apred}}) \iff (\struc,\store) \models
  \charform{r}{\tree}{\apred(\xi_1, \ldots, \xi_{\arityof{\apred}})}$,
  where $\tree=(\tnodes,\tedges,r,\alabel)$ is an unfolding tree for
  $\apred(\xi_1, \ldots, \xi_{\arityof{\apred}})$.
\end{lemmaE}
\begin{proofE}
  (We omit the annotations $\cdot^n$, as in $\phi^n$, in this proof and simply write $\phi$ because the annotations are not relevant for this proof.)

  ``$\Rightarrow$'' By induction on the definition of the satisfaction
  relation $(\struc,\store) \models_\asid \apred(\xi_1, \ldots,
  \xi_{\arityof{\apred}})$. Assume the relation holds because:
  \[(\struc,\store') \models_\asid \psi\overline{s} *
  \Asterisk_{\ell=1}^h \bpred_\ell(z_{\ell,1}, \ldots,
  z_{\ell,\arityof{\bpred_\ell}})\overline{s}\] for a rule $\arule :
  \apred(x_1, \ldots, x_{\arityof{\apred}}) \leftarrow \exists y_1
  \ldots \exists y_m ~.~ \psi * \Asterisk_{\ell=1}^h
  \bpred_\ell(z_{\ell,1}, \ldots, z_{\ell,\arityof{\bpred_\ell}})$
  from $\asid$, where $\psi$ is a quantifier- and predicate-free
  formula, $\overline{s} \isdef
  [x_1/\xi_1,\ldots,x_{\arityof{\apred}}/\xi_{\arityof{\apred}}]$ is a
  substitution and $\store'$ is a store that agrees with $\store$ over
  $\set{\xi_1, \ldots, \xi_{\arityof{\apred}}} \cap \vars$. Then there
  exist structures $\struc_0 \comp \struc_1 \comp \ldots \comp
  \struc_h = \struc$, such that $(\struc_0,\store') \models
  \psi\overline{s}$ and $(\struc_\ell,\store') \models_\asid
  \bpred_\ell(z_{\ell,1}, \ldots,
  z_{\ell,\arityof{\bpred_\ell}})\overline{s}$, for all $\ell \in
  \interv{1}{h}$. By the induction hypothesis, there exist unfolding
  trees $\tree_\ell = (\tnodes_\ell,\tedges_\ell,r_\ell,\alabel_\ell)$
  for $\bpred_\ell(z_{\ell,1}, \ldots,
  z_{\ell,\arityof{\bpred_\ell}})\overline{s}$, such that
  $(\struc_\ell,\store') \models_\asid
  \charform{r_\ell}{\tree_\ell}{\bpred_\ell(z_{\ell,1}, \ldots,
    z_{\ell,\arityof{\bpred_\ell}})\overline{s}}$, for all $\ell \in
  \interv{1}{h}$. Then $\tree = (\tnodes,\tedges,r,\alabel)$ is
  defined as $\tnodes \isdef \set{r} \cup \bigcup_{\ell=1}^h
  \tnodes_\ell$, $\tedges \isdef \set{(r,r_\ell) \mid \ell \in
    \interv{1}{h}} \cup \bigcup_{\ell=1}^h \tedges_\ell$ and $\alabel
  = \set{(r,\arule)} \cup \bigcup_{\ell=1}^h \alabel_\ell$, assuming
  w.l.o.g. that $\tnodes_\ell \cap \tnodes_k = \emptyset$, for all $1
  \leq \ell < k \leq h$ and $r \not\in
  \bigcup_{\ell=1}^h\tnodes_\ell$. The check that $(\struc,\store)
  \models \charform{r}{\tree}{\apred(\xi_1, \ldots,
    \xi_{\arityof{\apred}})}$ is routine.

  \noindent''$\Leftarrow$'' By induction on the structure of
  $\tree$. Let $p_1, \ldots, p_h$ be the children of $r$ in
  $\tree$. By Def. \ref{def:unfolding-tree}, we have:
  \[(\struc,\store') \models \psi\overline{s} *
  \Asterisk_{\ell=1}^h
  \charform{p_\ell}{\proj{\tree}{p_\ell}}{\bpred_\ell(z_{\ell,1},
    \ldots, z_{\ell,\arityof{\bpred_\ell}})}\overline{s}\] where
  $\overline{s}\isdef[x_1/\xi_1,\ldots,x_{\arityof{\apred}}/\xi_{\arityof{\apred}}]$
  is a substitution, $\store'$ is a store that agrees with $\store$
  over $\set{\xi_1, \ldots, \xi_{\arityof{\apred}}} \cap \vars$ and
  $\alabel(r)=\apred(x_1, \ldots, x_{\arityof{\apred}}) \leftarrow
  \exists y_1 \ldots \exists y_m ~.~ \psi * \Asterisk_{\ell=1}^h
  \bpred_\ell(z_{\ell,1}, \ldots, z_{\ell,\arityof{\bpred_\ell}})$ is
  a rule from $\asid$. Then there exist structures $\struc_0 \comp
  \struc_1 \comp \ldots \comp \struc_h = \struc$, such that
  $(\struc_0,\store') \models \psi\overline{s}$ and
  $(\struc_\ell,\store') \models
  \charform{p_\ell}{\proj{\tree}{p_\ell}}{\bpred_\ell(z_{\ell,1},
    \ldots, z_{\ell,\arityof{\bpred_\ell}})}\overline{s}$, for all
  $\ell \in \interv{1}{h}$. Since $\proj{\tree}{p_\ell}$ is an
  unfolding tree for $\bpred_\ell(z_{\ell,1}, \ldots,
  z_{\ell,\arityof{\bpred_\ell}})\overline{s}$, by the inductive
  hypothesis, we obtain $(\struc_\ell,\store') \models_\asid
  \bpred_\ell(z_{\ell,1}, \ldots,
  z_{\ell,\arityof{\bpred_\ell}})\overline{s}$, for all $\ell \in
  \interv{1}{h}$. Then, we have $(\struc,\store') \models_\asid
  \psi\overline{s} * \Asterisk_{\ell=1}^h \bpred_\ell(z_{\ell,1},
  \ldots, z_{\ell,\arityof{\bpred_\ell}})\overline{s}$, leading to
  $(\struc,\store) \models_\asid \apred(\xi_1, \ldots,
  \xi_{\arityof{\apred}})$.
\end{proofE}

To build a \sol\ formula that defines the models of a \slr\ sentence
$\phi$, we first add a rule $\apred_\phi() \leftarrow \phi$ to $\asid$
and describe a formula that asserts the existence of an unfolding tree
for $\apred_\phi()$ in the SID $\asid_\phi \isdef \asid \cup
\set{\apred_\phi() \leftarrow \phi}$. Let $\npred$ be the maximum
number of occurrences of predicate atoms in a rule from $\asid_\phi$.
We use second-order variables $Y_1, \ldots, Y_\npred$ of arity $2$,
for the edges of the tree and $X_1, \ldots, X_\nrule$ of arity $1$,
for the labels of the nodes in the tree i.e., the rules of $\asid$.
Then the \sol\ formula $\mathfrak{T}(x,
\set{X_i}_{i=1}^\nrule,\set{Y_j}_{j=1}^\npred)$ is defined as the
conjunction of \sol\ formul{\ae} that describe the following facts:
\begin{compactenum}
\item the root $x$ belongs to $X_i$, for some rule $\arule_i$ that
  defines the predicate $\apred_\phi$,
\item the sets $X_1, \ldots, X_\nrule$ are pairwise disjoint,
\item each vertex in $X_1 \cup \ldots \cup X_\nrule$ is reachable from $x$
  by a path of edges $Y_1, \ldots, Y_\npred$,
\item each vertex in $X_1 \cup \ldots \cup X_\nrule$, except for $x$, has
  exactly one incoming edge and $x$ has no incoming edge,
\item each vertex from $X_i$ has exactly $h$ outgoing edges $Y_1,
  \ldots, Y_h$, each to a vertex from $X_{j_\ell}$, respectively, such
  that $\arule_{j_\ell}$ defines the predicate $\bpred_\ell$, for all
  $\ell \in \interv{1}{h}$, where $\bpred_1(z_{1,1}, \ldots,
  z_{1,\arityof{\bpred_1}}), \ldots, \bpred_h(z_{h,1}, \ldots,
    z_{h,\arityof{\bpred_h}})$ are the predicate atoms that occur in
    $\arule_i$.
\end{compactenum}
We now state a \sol\ formula $\mathfrak{F}(\xi_1, \ldots,
\xi_{\arityof{\apred}}, x, \set{X_i}_{i=1}^\nrule,
\set{Y_j}_{j=1}^\npred,
\set{\set{Z_{k,\ell}}_{\ell=1}^{\arityof{\arel_k}}}_{k=1}^\nrel)$ that
fixes the relationship between the unfolding tree $\tree$ and the
relations $\struc(\arel_i)$.  We recall that by construction for every
node $n$ of $\tree$ and every relation atom $\arel$ there is at most
one annotated relation atom $\arel_k^n(\xi_1,
\ldots,\xi_{\arityof{\arel_i}})$ in
$\charform{r}{\tree}{\apred(\xi_1,\ldots,\xi_{\arityof{\apred}})}$.
The formula $\mathfrak{F}$ now uses second-order variables
$Z_{k,\ell}$, of arity $2$, in order to encode (partial) functions
mapping a tree vertex $n$ to the value of $\xi_\ell$ for the (unique)
annotated atom $\arel_k^n(\xi_1, \ldots,\xi_{\arityof{\arel_i}})$ (in
case such an atom exists).  The formula $\mathfrak{F}$ is the
conjunction of following \sol-definable
facts: \begin{compactenum}[(i)]
\item\label{it1:slr-so} each second-order variable $Z_{k,\ell}$
  denotes a functional binary relation:
  \vspace*{-.5\baselineskip}
  \[\bigwedge_{k\in\interv{1}{\nrel}} \bigwedge_{\ell \in \interv{1}{\arityof{\arel_k}}}
  \forall y \forall z \forall z' ~.~ Z_{k,\ell}(y,z) \wedge Z_{k,\ell}(y,z') \rightarrow z=z'\]
  \vspace*{-.5\baselineskip}
\item\label{it2:slr-so} for each relation atom with relation symbol
  $\arel_k$ that occurs in a rule $\arule_i$, the $\arel_k$-relation
  contains a tuple:
  \vspace*{-.5\baselineskip}
  \[\begin{array}{l}
  \bigwedge_{i \in \interv{1}{\nrule}} \bigwedge_{\arel_k \text{
      occurs in } \arule_i} \forall y ~.~ X_i(y) \rightarrow \exists
  z_1 \ldots \exists z_{\arityof{\arel_k}} ~.~ \arel_k(z_1,\ldots,
  z_{\arityof{\arel_k}}) \wedge \bigwedge_{\ell \in
    \interv{1}{\arityof{\arel_k}}} Z_{k,\ell}(y,z_\ell)
  \end{array}\]
  \vspace*{-.5\baselineskip}
\item\label{it3:slr-so} for any two (not necessarily distinct) rules $\arule_i$ and
  $\arule_j$ such that an atom with relation symbol $\arel_k$ occurs
  in both, the tuples introduced by the two atoms are distinct:
  \vspace*{-.5\baselineskip}
  \[\begin{array}{l}
  \bigwedge_{i, j \in \interv{1}{\nrule}} \bigwedge_{\arel_k \text{ occurs in } \arule_i, \arule_j}
  \forall y \forall y' \forall z_1 \forall z'_1 \ldots \forall z_{\arityof{\arel_k}} \forall z'_{\arityof{\arel_k}} ~.~ \\[2mm]
  \Big(X_i(y) \wedge X_j(y') \wedge \bigwedge_{\ell\in\interv{1}{\arityof{\arel_k}}} (Z_{k,\ell}(y,z_\ell) \wedge Z_{k,\ell}(y',z'_\ell))\Big)
  \rightarrow \bigvee_{\ell\in\interv{1}{\arityof{\arel_k}}} z_\ell \neq z'_\ell
  \end{array}\]
  \vspace*{-.5\baselineskip}
\item\label{it4:slr-so} each tuple from a $\arel_k$-relation must have been introduced
  by a relation atom with relation symbol $\arel_k$ that occurs in a rule
  $\arule_i$:
  \vspace*{-.5\baselineskip}
  \[\bigwedge_{k \in \interv{1}{\nrel}}\forall z_1 \ldots \forall z_{\arityof{\arel_k}} ~.~ \arel_k(z_1, \ldots, z_{\arityof{\arel_k}}) \rightarrow
  \exists y ~.~ \hspace*{-4mm} \bigvee_{\arel_k \text{ occurs in } \arule_i} \hspace*{-2mm} \Big(X_i(y) ~\wedge
  \bigwedge_{\ell \in \interv{1}{\arityof{\arel_k}}} Z_{k,\ell}(y,z_\ell)\Big)\]
  \vspace*{-.5\baselineskip}
\item\label{it5:slr-so} two terms $\xi_m$ and $\chi_n$ that occur
  within two relation atoms $\arel_k(\xi_1, \ldots,
  \xi_{\arityof{\arel_k}})$ and $\arel_\ell(\chi_1, \ldots,
  \chi_{\arityof{\arel_\ell}})$ within rules $\arule_i$ and
  $\arule_j$, respectively, and are constrained to be equal (i.e., via
  equalities and parameter passing), must be equated:
  \vspace*{-.5\baselineskip}
  \[\begin{array}{l}
  \bigwedge_{k,\ell \in \interv{1}{\nrel}} \bigwedge_{\begin{array}{l}
      \scriptstyle{\arel_k \text{ occurs in } \arule_i} \\[-1mm]
      \scriptstyle{\arel_\ell \text{ occurs in } \arule_j}
  \end{array}} \bigwedge_{m \in \interv{1}{\arityof{\arel_k}}} \bigwedge_{n \in \interv{1}{\arityof{\arel_\ell}}}
  \forall y \forall y' \forall z' \forall z' ~.~ \\
  \Big(X_i(y) \wedge X_j(y') \wedge \mathit{isEq}_{k,\ell,m,n}(y, y', \set{X_i}_{i=1}^\nrule,
  \set{Y_j}_{j=1}^\npred) \wedge Z_{k,m}(y,z) \wedge Z_{\ell,n}(y',z')\Big)
  \rightarrow z = z'
  \end{array}\]

  \vspace*{-.5\baselineskip}\noindent The formula
  $\mathit{isEq}_{k,\ell,m,n}(y,y',\set{X_i}_{i=1}^\nrule,\set{Y_j}_{j=1}^\npred)$
  asserts that there is a path in the unfolding tree between the store
  values (i.e., vertices of the tree) of $y$ and $y'$, such that the
  $m$-th and $n$-th variables of the relation atoms $\arel_k(z_1,
  \ldots, z_{\arityof{\arel_k}})$ and $\arel_\ell(z'_1, \ldots,
  z'_{\arityof{\arel_\ell}})$ are bound to the same value.
  %
\item\label{it6:slr-so} a disequality $\xi \neq \chi$ that occurs in a rule $\arule_i$
  is propagated throughout the tree to each pair of variables that
  occur within two relation atoms $\arel_k(\xi_1, \ldots,
  \xi_{\arityof{\arel_k}})$ and $\arel_\ell(\chi_1, \ldots,
  \chi_{\arityof{\arel_\ell}})$ in rules $\arule_{j_k}$ and
  $\arule_{j_\ell}$, respectively, such that $\xi$ is bound $\xi_r$ and
  $\chi$ to $\chi_s$ by equality atoms and parameter passing:
  \vspace*{-.5\baselineskip}
  \[\begin{array}{l}
  \bigwedge_{\xi \neq \chi \text{ occurs in } \arule_i}
  \bigwedge_{k,\ell \in \interv{1}{\nrel}} \bigwedge_{\begin{array}{l}
      \scriptstyle{\arel_{k} \text{ occurs in } \arule_{j_k}} \\[-1mm]
      \scriptstyle{\arel_{\ell} \text{ occurs in } \arule_{j_\ell}}
  \end{array}} \bigwedge_{r \in \interv{1}{\arityof{\arel_k}}} \bigwedge_{s\in\interv{1}{\arityof{\arel_\ell}}}
  \forall y \forall y' \forall y'' \forall z' \forall z'' ~.~ \\
  \Big(X_i(y) \wedge X_{j_k}(y') \wedge X_{j_\ell}(y'') \wedge Z_{k,r}(y',z') \wedge Z_{\ell,s}(y'',z'') \\
  \hspace*{4mm} \mathit{varEq}_{\xi,k,r}(y, y', \set{X_i}_{i=1}^N, \set{Y_j}_{j=1}^M) \wedge
  \mathit{varEq}_{\chi,\ell,s}(y, y'', \set{X_i}_{i=1}^N) \Big) \rightarrow z' \neq z''
  \end{array}\]
  The formula
  $\mathit{varEq}_{\xi,k,r}(x,y,\set{X_i}_{i=1}^\nrule,\set{Y_j}_{j=1}^\npred)$
  states that the variable $\xi$ occuring in the label of the
  unfolding tree vertex $x$ is bound to the variable $z_r$ that occurs
  in a relation atom $\arel_k(z_1, \ldots, z_{\arityof{\arel_i}})$ in
  the label of the vertex $y$.
\item\label{it7:slr-so} each top-level variable in $\apred(\xi_1,
  \ldots, \xi_{\arityof{\apred}})$ that is bound to a variable from a
  relation atom $\arel_k(z_1, \ldots, z_{\arityof{\arel_k}})$ in the
  unfolding, must be equated to that variable:
  \vspace*{-.5\baselineskip}
  \[\bigwedge_{\arule_i \in \defn{\asid}{\apred}} \bigwedge_{j \in \interv{1}{\arityof{\apred}}} \forall y \forall z ~.~
  X_i(x) \wedge \mathit{varEq}_{\xi_j,k,r}(x, y, \set{X_i}_{i=1}^\nrule,\set{Y_j}_{j=1}^\npred) \wedge Z_{k,r}(y,z) \rightarrow \xi_j = z\]
\end{compactenum}
The formul{\ae}
$\mathit{isEq}_{k,\ell,m,n}(x,y,\set{X_i}_{i=1}^\nrule,\set{Y_j}_{j=1}^\npred)$
and
$\mathit{varEq}_{\xi,k,r}(x,y,\set{X_i}_{i=1}^\nrule,\set{Y_j}_{j=1}^\npred)$
above are definable in \mso, using standard tree automata construction
techniques, similar to the definition of \mso\ formul{\ae} that track
parameters in an unfolding tree for \seplog, with edges definable by
\mso\ formul{\ae} over the signature of
\seplog\ \cite{DBLP:conf/cade/IosifRS13}. To avoid clutter, we defer
such definitions to a long version of this paper.

Summing up, the \sol\ formula defining the models of the predicate
atom $\apred(\xi_1, \ldots, \xi_{\arityof{\apred}})$ with respect to
the SID $\asid$ is the following:
\vspace*{-.5\baselineskip}
\begin{align*}
\mathfrak{A}^\apred_\asid(\xi_1, \ldots, \xi_{\arityof{\apred}}) ~\isdef &
~\exists x \exists \set{X_i}_{i=1}^\nrule \exists \set{Y_j}_{j=1}^\npred
\exists \set{Z_{1,\ell}}_{\ell=1}^{\arityof{\arel_1}} \ldots \exists
\set{Z_{K,\ell}}_{\ell=1}^{\arityof{\arel_K}} ~.~ \\
& \mathfrak{T}(x,\set{X_i}_{i=1}^\nrule,\set{Y_j}_{j=1}^\npred) \wedge
\mathfrak{F}(\xi_1, \ldots, \xi_{\arityof{\apred}},x, \set{X_i}_{i=1}^\nrule, \set{Y_j}_{j=1}^\npred,
\set{\set{Z_{k,\ell}}_{\ell=1}^{\arityof{\arel_k}}}_{k=1}^\nrel)
\end{align*}
The correctness of the construction is asserted by the following
proposition, that also proves \slr\ $\subseteq$\ \sol, using a similar
reduction of sentences to predicate atoms as in Proposition
\ref{prop:slr-not-strictly-unbounded}:

\begin{propositionE}\label{prop:slr-so}
  Given an SID $\asid$ and a predicate atom $\apred(\xi_1, \ldots,
  \xi_{\arityof{\apred}})$, for each structure $\struc$ and store
  $\store$, we have $(\struc,\store) \models_{\asid} \apred(\xi_1,
  \ldots, \xi_{\arityof{\apred}}) \iff (\struc,\store) \Models
  \mathfrak{A}^\apred_\asid(\xi_1, \ldots, \xi_{\arityof{\apred}})$.
\end{propositionE}
\begin{proofE}
  ``$\Rightarrow$'' By Lemma \ref{lemma:unfolding-tree}, there exists
  an unfolding tree $\tree = (\tnodes,\tedges,r,\alabel)$ of
  $\apred(\xi_1, \ldots, \xi_{\arityof{\apred}})$, such that
  $(\struc,\store) \models \charform{r}{\tree}{\apred(\xi_1, \ldots,
    \xi_{\arityof{\apred}})}$.  Let $\charform{r}{\tree}{\apred(\xi_1,
    \ldots, \xi_{\arityof{\apred}})} = \exists y_1 \ldots \exists y_K
  ~.~ \Phi$, where $\Phi$ is a quantifier- and predicate-free formula.
  Note that, by Def. \ref{def:unfolding-tree}, no second-order
  variables occur in $\charform{r}{\tree}{\apred(\xi_1, \ldots,
    \xi_{\arityof{\apred}})}$. Hence there exists a store $\store'$
  that agrees with $\store$ over $\xi_1, \ldots,
  \xi_{\arityof{\apred}}$, such that $(\struc,\store') \models
  \Phi$. We define another store $\store''$, that agrees with $\store$
  and $\store'$ over $\xi_1, \ldots, \xi_{\arityof{\apred}}$ such
  that, moreover, we have: \begin{compactitem}
  \item $\store''(x)=r$,
  \item $\store''(X_i) = \set{n \in \tnodes \mid \alabel(n) =
    \arule_i}$, for all $i \in \interv{1}{\nrule}$,
  \item $\store''(Y_j) = \set{(n,m) \in \tnodes \times \tnodes \mid m
    \text{ is the $j$-th child of } n}$, for all $j \in
    \interv{1}{\npred}$; we consider that the order between the
    children of a vertex in an unfolding tree is the syntactic order
    of their corresponding predicate atoms, in the sense of
    Def. \ref{def:unfolding-tree},
  \item $\store''(Z_{k,\ell}) = \set{(n,(\struc,\store')(\xi_\ell)) \mid n \in
    \tnodes,~ \arel^n_k(\xi_1,\ldots,\xi_{\arityof{\arel_k}}) \text{
      occurs in } \Phi}$, for all $k \in \interv{1}{\nrel}$ and $\ell
    \in \interv{1}{\arityof{\arel_k}}$.
  \end{compactitem}
  We have $(\struc,\store'') \Models \mathfrak{T}(x,
  \set{X_i}_{i=1}^\nrule, \set{Y_j}_{j=1}^\npred)$ because $\tree$ is
  an unfolding tree for $\apred(\xi_1,\ldots,\xi_{\arityof{\apred}})$,
  by Def. \ref{def:unfolding-tree}. The proof of \((\struc,\store'')
  \Models \mathfrak{F}(\xi_1, \ldots, \xi_{\arityof{\apred}},x,
  \set{X_i}_{i=1}^\nrule, \set{Y_j}_{j=1}^\npred,
  \set{\set{Z_{k,\ell}}_{\ell=1}^{\arityof{\arel_k}}}_{k=1}^\nrel)\)
  follows from $(\struc,\store') \models \Phi$ and the definition of
  $\store''$, by the points (\ref{it1:slr-so}-\ref{it7:slr-so}) from
  the definition of $\mathfrak{F}$. We obtain $(\struc,\store) \Models
  \mathfrak{A}^\apred_\asid(\xi_1, \ldots, \xi_{\arityof{\apred}})$
  from the definition of $\mathfrak{A}^\apred_\asid$.

  \noindent''$\Leftarrow$'' There exists a store $\store'$ that agrees
  with $\store$ over $\xi_1, \ldots, \xi_{\arityof{\apred}}$, such
  that:
  \begin{align}
  (\struc,\store') \Models & \mathfrak{T}(x, \set{X_i}_{i=1}^\nrule,
    \set{Y_j}_{j=1}^\npred) \label{eq:tree} \\
    (\struc,\store') \Models &
    \mathfrak{F}(\xi_1, \ldots, \xi_{\arityof{\apred}},x,
    \set{X_i}_{i=1}^\nrule, \set{Y_j}_{j=1}^\npred,
    \set{\set{Z_{k,\ell}}_{\ell=1}^{\arityof{\arel_k}}}_{k=1}^\nrel) \label{eq:form}
  \end{align}
  By (\ref{eq:tree}) we obtain an unfolding tree $\tree =
  (\tnodes,\tedges,r,\alabel)$ for $\apred(\xi_1, \ldots,
  \xi_{\arityof{\apred}})$, such that: \begin{compactitem}
  \item $\tnodes = \bigcup_{i=1}^\nrule \store'(X_i)$,
  \item $\tedges = \bigcup_{j=1}^\npred \store'(Y_j)$,
  \item $\alabel(n) = \arule_i \iff n \in \store'(X_i)$, for all $n
    \in \tnodes$ and $i \in \interv{1}{\nrule}$.
  \end{compactitem}
  Let $\charform{r}{\tree}{\apred(\xi_1, \ldots,
    \xi_{\arityof{\apred}})} = \exists y_1 \ldots \exists y_K ~.~
  \Phi$, where $\Phi$ is a quantifier- and predicate-free formula
  (Def. \ref{def:unfolding-tree}). By Lemma
  \ref{lemma:unfolding-tree}, it is sufficient to show the existence
  of a store $\store''$ that agrees with $\store$ over $\xi_1, \ldots,
  \xi_{\arityof{\apred}}$, such that $(\struc,\store'') \models
  \Phi$. Let $f_{k,\ell}$ denote the partial mapping defined by
  $\store'(Z_{k,\ell})$, for each $k \in \interv{1}{\nrel}$ and $\ell
  \in \interv{1}{\arityof{\arel}}$, by point (\ref{it1:slr-so}) of the
  definition of $\mathfrak{F}$. For each $r \in \interv{1}{K}$, we
  define $\store''(y_r) \isdef f_{k,\ell}(n)$ if $y_r$ occurs in or is
  constrained to be equal to a term $\xi_\ell$ that occurs in an
  annotated relation atom $\arel^n_k(\xi_1, \ldots,
  \xi_{\arityof{\arel_k}})$ from $\Phi$. Note that there can by at
  most one such relation atom in $\Phi$, because of the assumption
  that in each rule from $\asid$ at most one relation atom
  $\arel_k(z_1,\ldots,z_{\arityof{\arel_k}})$ occurs. Otherwise, if
  $y_r$ is not constrained in $\Phi$ to be equal to a term that occurs
  in a relation atom, $\store''(y_r)$ is given an arbitrary fresh
  value. Because the equalities and disequalities from $\Phi$ are
  taken care of by points (\ref{it5:slr-so}) and (\ref{it6:slr-so})
  from the definition of $\mathfrak{F}$, it remains to check the
  satisfaction of the relation atoms from $\Phi$. To this end, we
  define a decomposition $\struc = \bigcomp_{n \in \tnodes,~ k \in
    \interv{1}{N}} \struc_{n,k}$ such that $(\struc_{n,k},\store'')
  \models \arel_k(\xi_1,\ldots,\xi_{\arityof{\arel_k}})$, for all
  relation atoms $\arel_k^n(\xi_1,\ldots,\xi_{\arityof{\arel_k}})$
  from $\Phi$. Such a decomposition is possible due to points
  (\ref{it2:slr-so}-\ref{it4:slr-so}) from the definition of
  $\mathfrak{F}$.
\end{proofE}

\section{\mso\ $\subseteq^k$ \slr}
\label{sec:k-mso-slr}

We prove that, for any \mso\ sentence $\phi$ and any integer $k \geq
1$, there exists an SID $\twformsid{k}{\phi}$ that defines a predicate
$\apred_{k,\phi}$ of arity zero, such that the set of models of $\phi$
of treewidth at most $k$ corresponds to the set of
structures \slr-defined by the pair
$(\apred_{k,\phi}(),\twformsid{k}{\phi})$. Our proof leverages from a
result of Courcelle
\cite{DBLP:journals/tcs/Courcelle92}, stating that the set of models of
bounded treewidth of a given \mso\ sentence can be described by a set
of recursive equations, written using an algebra of operations on
structures. This result follows up in a long-standing line of work
(known as Feferman-Vaught theorems
\cite{journals/apal/Makowsky04}) that reduces the evaluation of an
\mso\ sentence on the result of an algebraic operation to the evaluation
of several related sentences in the arguments of the respective
operation.

\vspace*{-.5\baselineskip}
\subsection{Courcelle's Theorem}
\label{sec:courcelle}

In order to exlpain our construction (given
in \S\ref{sec:encoding-types-sids}), we recall first a result of
Courcelle on the characterization of the structures of bounded
treewidth that satisfy a given MSO formula $\phi$ by an effectively
constructible set of recursive equations.  The equation set uses two
operations on structures, namely $\glue$ and $\fgcst{j}$, that are
lifted to sets of structures, as usual. The result is developed in two
steps: the first step builds a generic equation set, that
characterizes all structures of bounded treewidth, which is then
refined to describe only those structures that satisfy $\phi$, in the
second step.

\paragraph{Standard Structures}
Given a signature $\signature$, a \emph{standard
  structure} $\astruc = (\adomof{\astruc},\interpof{\astruc})$
consists of a domain $\adomof{\astruc}$ and an interpretation
$\interpof{\astruc}$ of the relation and constant symbols from
$\signature$ as tuples and elements from $\adomof{\astruc}$,
respectively. Note that the structures $\astruc$ considered in the
semantics of \slr\ correspond to the particular case
$\adomof{\astruc}=\universe$, for a countably infinite set
$\universe$, whereas in general, $\adomof{\astruc}$ might be finite or
even uncountable. The interpretation of \solmso\ formul{\ae} and the
notion of treewidth are defined for standard structures in the same
way as for structures $(\universe,\struc)$. We denote by
$\strucof{\signature}$ the set of standard structures over the
signature $\signature$.
We sometimes abuse notation and write $\struc \in
\strucof{\signature}$ instead of $(\universe,\struc) \in
\strucof{\signature}$.

\paragraph{Operations}
Let $\signature
= \set{\arel_1, \ldots, \arel_\nrel, \acst_1, \ldots, \acst_\ncst}$
and $\signature'
= \set{\arel'_1, \ldots, \arel'_{\nrel'}, \acst'_1, \ldots, \acst'_{\ncst'}}$
be two (possibly overlapping) signatures.  The \emph{glueing}
operation $\glue
: \strucof{\signature} \times \strucof{\signature'} \rightarrow
\strucof{\signature\cup\signature'}$
is the disjoint union of structures, followed by the fusion of
constants. Formally, given $\astruc_i = (\adomof{i},\interpof{i})$, for
$i = 1,2$, such that $\adomof{1} \cap \adomof{2} = \emptyset$, let
$\sim$ be the least equivalence relation on $\adomof{1} \cup
\adomof{2}$ such that $\interpof{1}(\acst) \sim \interpof{2}(\acst)$,
for all $\acst \in \signature \cap \signature'$. We denote by $[d]$
the equivalence class of an element $d \in \adomof{1} \cup
\adomof{2}$ w.r.t. the $\sim$ relation. Then $\glueof{\astruc_1}{\astruc_2} \isdef (\adomof{},
\interpof{})$, where $\adomof{} \isdef \set{[d] \mid d \in \adomof{1}
  \cup \adomof{2}}$ and the interpretation of relation symbols and
constants follows:
\vspace*{-.5\baselineskip}
\[\begin{array}{rcl}
\interpof{}(\arel) & \isdef & \left\{\begin{array}{ll}
\set{\tuple{[d_1], \ldots [d_{\arityof{\arel}}]} \mid \tuple{d_1, \ldots, d_{\arityof{\arel}}} \in \interpof{1}(\arel)}, &
\text{if } \arel \in \signature\setminus\signature' \\
\set{\tuple{[d_1], \ldots [d_{\arityof{\arel}}]} \mid \tuple{d_1, \ldots, d_{\arityof{\arel}}} \in \interpof{2}(\arel)}, &
\text{if } \arel \in \signature'\setminus\signature \\
\set{\tuple{[d_1], \ldots [d_{\arityof{\arel}}]} \mid \tuple{d_1, \ldots, d_{\arityof{\arel}}} \in \interpof{1}(\arel) \cup \interpof{2}(\arel)}, &
\text{if } \arel \in \signature \cap \signature
\end{array}\right. \\[2mm]
\interpof{}(\acst) & \isdef & \left\{\begin{array}{ll}
\text{$[\interpof{1}(\acst)]$}, & \text{if } \acst \in \signature \\
\text{$[\interpof{2}(\acst)]$}, & \text{if } \acst \in \signature' \setminus \signature
\end{array}\right.
\end{array}\]

\vspace*{-.5\baselineskip}\noindent
Since we identify isomorphic structures, the nature of the elements of
$\adomof{}$ (i.e., equivalence classes of the $\sim$ relation) is not
important. The \emph{forget} operation $\fgcst{j} :
\strucof{\signature} \rightarrow \strucof{\signature \setminus
\set{\acst_j}}$ simply drops the constant $\acst_j$ from the domain of its argument,
for any $j \in \interv{1}{\ncst}$.

\paragraph{Structures of Bounded Treewidth}
Let $k\geq1$ be an integer, $\signature
= \{\arel_1, \ldots, \arel_\nrel, \acst_1, \ldots, \acst_\ncst\}$ be a
signature and $\ports
= \set{\acst_{\ncst+1},\ldots,\acst_{\ncst+k+1}}$ be a set of
constants disjoint from $\signature$, called \emph{ports}. We consider
variables $Y_i$, for all subsets $\ports_i \subseteq \ports$, denoting
sets of structures over the signature $\signature \cup \ports_i$. The
equation system now consists of recursive equations of the form
$Y_0 \supseteq f(Y_1, \ldots, Y_n)$, where each $f$ is either $\glue$,
$\fgcst{\ncst + j}$, for any $j \in \interv{1}{k+1}$, or a relation
$\overline{\arel}_i$ of type $\arel_i$, consisting of one tuple with
at most $k+1$ distinct elements, for any $i \in \interv{1}{\nrel}$. We
denote this set of equations by $\tweq{k}$. The structures of
treewidth at most $k$ correspond to a component of the least solution
of $\tweq{k}$, in the domain of tuples of sets ordered by pointwise
inclusion, see e.g., \cite[Theorem 2.83]{courcelle_engelfriet_2012}.

\paragraph{Refinement to Models of \mso}
We recall that the \emph{quantifier rank} $\qrof{\phi}$ of an
\mso\ formula $\phi$ is the maximal depth of nested quantifiers i.e.,
$\qrof{\phi} \isdef 0$ if $\phi$ is an atom, $\qrof{\neg\phi_1} \isdef
\qrof{\phi_1}$, $\qrof{\phi_1 \wedge \phi_2} \isdef
\max(\qrof{\phi_1}, \qrof{\phi_2})$ and $\qrof{\exists x ~.~ \phi_1} =
\qrof{\exists X ~.~ \phi_1} \isdef \qrof{\phi_1} + 1$. We denote by
$\typesof{r}$ the set of \mso\ sentences of quantifier rank at most
$r$; this set is finite, up to logical equivalence. For a standard
structure $\astruc$, we define its $r$-\emph{type} as
$\typeof{r}{\astruc} \isdef \set{\phi \in \typesof{r} \mid \astruc \Models \phi}$. We
assume the sentences in $\typeof{r}{\astruc}$ to use the signature
over which $\astruc$ is defined; this signature will be clear from the
context in the following.

\begin{definition}\label{def:compatible}
An operation $f : \strucof{\signature_1} \times \ldots \times
\strucof{\signature_n} \rightarrow \strucof{\signature_{n+1}}$ is said
to be (effectively) \mso-\emph{compatible}\footnote{Also referred to
  as \emph{smooth} in \cite{journals/apal/Makowsky04}.} iff
  $\typeof{r}{f(\astruc_1, \ldots, \astruc_n)}$ depends only on (and
  can be computed from) $\typeof{r}{\astruc_1},
\ldots, \typeof{r}{\astruc_n}$ by an \emph{abstract operation}
$\absof{f}
: \left(\pow{\typesof{r}}\right)^n \rightarrow \pow{\typesof{r}}$.
\end{definition}

Courcelle establishes that glueing and forgetting of constants are
effectively \mso-compatible and computable by abstract operations
$\absglue$ and $\absfgcst{\ncst + i}$, for $i \in
\interv{1}{k+1}$. Then, one can build, from $\tweq{k}$ a set of recursive equations $\abstweq{k}$ of the
form $Y_0^{\tau_0} = f(Y_1^{\tau_1}, \ldots, Y_n^{\tau_n})$, where
$Y_0 = f(Y_1, \ldots, Y_n)$ is an equation from $\tweq{k}$ and
$\tau_0, \ldots, \tau_n$ are $\qrof{\phi}$-types such that $\tau_0 =
\absof{f}(\tau_1, \ldots, \tau_n)$.
Intuitively, each annotated variable $Y^\tau$ denotes denotes the
set of structures whose $\qrof{\phi}$-type is $\tau$, from the
$Y$-component of the the least solution of $\tweq{k}$. Then, the set
of models of $\phi$ of treewidth at most $k$ is the union of the
$Y^\tau$-components of the least solution of $\abstweq{k}$, such that
$\phi \in \tau$, see e.g., \cite[Theorem 3.6]{DBLP:journals/tcs/Courcelle92}.

\subsection{Encoding Types with SIDs}
\label{sec:encoding-types-sids}

We now give the details of \mso\ $\subseteq^k$\ \slr. Instead of using
the set of recursive equations $\tweq{k}$ from the previous
subsection, we give an SID $\twsid{k}$ that also characterizes the
structures of bounded treewidth (Figure \ref{fig:sids}a). The idea is
to use the separating conjunction for simulating the glueing
operation. However, the separating conjunction does
not necessarily guarantee disjointness of the structures, as required by the glueing operation. In order to enforce
disjointness with the separating conjunction, we introduce a new
relation symbol $\domsymb \not\in \signature$, that ``collects'' the values assigned to the existentially quantified
variables created by rule (\ref{rule:gen-exists}) and the top-level
rule (\ref{rule:gen-top}) during the unraveling. In particular, the
relation symbol $\domsymb$ ensures that
\begin{inparaenum}[(i)]
\item the variables of a predicate atom $\apred(x_1, \ldots,
  x_{\arityof{\apred}})$ are mapped to pairwise distinct values and
\item the composition $\struc_1 \comp \struc_2$ of two structures $\struc_1$ and $\struc_2$ over $\signature \cup \set{\domsymb}$ behaves like glueing the standard
  structures $(\Dom{\struc_1}, \struc_1)$ and $(\Dom{\struc_2}, \struc_2)$.
\end{inparaenum}
Formally, we establish a correspondence between structures over
$\signature$ and $\signature \cup \set{\domsymb}$: given a structure
$\struc \in \strucof{\signature}$, we say that
$\struc' \in \strucof{\signature \cup \set{\domsymb}}$ is
a \emph{$\domsymb$-extension} of $\struc$ if and only if
$\struc'(\domsymb) \supseteq \Rel{\struc}$ and $\struc'$ agrees with
$\struc$ over $\signature$.
%
%
Below we prove that $\twsid{k}$ (Figure \ref{fig:sids}a) fulfills the
purpose of defining the structures of bounded treewidth:


\begin{figure}[t!]
  \vspace*{-\baselineskip}
  \begin{align}
    \apred(x_1, \ldots, x_{k+1}) \leftarrow & ~\apred(x_1, \ldots,
    x_{k+1}) * \apred(x_1, \ldots, x_{k+1}) \label{rule:gen-comp}
    \\ \apred(x_1, \ldots, x_{k+1}) \leftarrow & ~\exists y ~.~
    \domsymb(y) * \apred(x_1, \ldots,x_{k+1})[x_i/y] \text{, for all }
    i \in \interv{1}{k+1} \label{rule:gen-exists} \\ \apred(x_1,
    \ldots, x_{k+1}) \leftarrow &
    ~\arel(y_1,\ldots,y_{\arityof{\arel}}), \text{for all }
    \arel\in\signature \text{ and } y_1,\ldots,y_{\arityof{\arel}} \in \set{x_1,
      \ldots, x_{k+1}} \label{rule:gen-rel}
    \\ \apred_k() \leftarrow & ~\exists x_1 \ldots \exists
    x_{k+1} ~.~ \domsymb(x_1) * \ldots * \domsymb(x_{k+1}) *
    \apred(x_1, \ldots, x_{k+1}) \label{rule:gen-top}
  \end{align}

  \vspace*{-.5\baselineskip}
  \centerline{\small(a)}

  \vspace*{-\baselineskip}
  \begin{align}
    \apred^\atype(x_1, \ldots, x_{k+1}) \leftarrow & ~\apred^{\atype_1}(x_1, \ldots, x_{k+1}) * \apred^{\atype_2}(x_1, \ldots, x_{k+1})
    \text{, where } \atype = \absglueof{\atype_1}{\atype_2} \label{rule:ref-comp} \\
    \apred^\atype(x_1, \ldots, x_{k+1}) \leftarrow & \exists y ~.~ \domsymb(y) * \apred^{\atype_1}(x_1, \ldots, x_{k+1})[x_i/y] \text{, for all } i \in \interv{1}{k+1},
    \label{rule:ref-exists} \\
    & \quad\quad \text{where } \atype = \absglueof{\absfgcst{\ncst+i}(\atype_1)}{\rho_i}
    \text{ for the type } \rho_i \nonumber \\
    & \quad\quad \text{of a structure }
    \astruc \in \strucof{\set{\acst_{\ncst+i}}} \text{ with singleton universe,} \nonumber \\
    \apred^\atype(x_1, \ldots, x_{k+1}) \leftarrow & ~\arel(y_1,\ldots,y_{\arityof{\arel}}) *
    \Asterisk_{\!\!\!\!\begin{array}{l} \scriptstyle{c \bowtie d \in \atype}, \\[-2mm] \scriptstyle{\bowtie \in \{=,\neq\}} \end{array}}
    \hspace*{-5mm} c \bowtie d [\acst_{\ncst+1}/x_1, \ldots, \acst_{\ncst+k+1}/x_{k+1}], \label{rule:ref-rel} \\
    & \quad\quad \text{where } \atype = \typeof{\qrof{\phi}}{\astruc},~ \astruc \in \strucof{\signature\cup\set{\acst_{\ncst+1},\ldots,\acst_{\ncst+k+1}}} \text{ and} \nonumber \\
    & \quad\quad \astruc \models \arel(y_1,\ldots,y_{\arityof{\arel}})[x_1/\acst_{\ncst+1}, \ldots, x_{k+1}/\acst_{\ncst+k+1}] \text{ for some } \nonumber \\
    & \quad\quad y_1, \ldots, y_{\arityof{\arel}} \in  \{x_1,\ldots,x_{k+1} \} \nonumber \\
    \apred_{k,\phi}() \leftarrow & ~\exists x_1 \ldots \exists x_{k+1} ~.~ \domsymb(x_1) * \ldots * \domsymb(x_{k+1}) *
    \apred^\atype(x_1, \ldots, x_{k+1}) \label{rule:ref-top} \\
    &  \quad\quad \text{for all } \atype \text{ such that } \phi\in\atype \nonumber
  \end{align}

  \vspace*{-.5\baselineskip}
  \centerline{\small(b)}
  \caption{The SID $\twsid{k}$ defining structures of treewidth at
    most $k$ (a) and its annotation $\twformsid{k}{\phi}$ defining the
    models of an \mso\ sentence $\phi$, of treewidth at most $k$ (b)}
  \label{fig:sids}
  \vspace*{-\baselineskip}
\end{figure}

\begin{myTextE}
\begin{definition}\label{def:reduced-td}
A tree decomposition $\tree=(\tnodes,\tedges,r,\alabel)$ of a
structure $\struc$ is said to be \emph{reduced} if and only if the
following hold: \begin{compactenum}
\item\label{it1:reduced-td} for each $\arel \in \signature$ and each $\tuple{u_1, \ldots,
  u_{\arityof{\arel}}} \in \struc(\arel)$ there exists a leaf $n \in
  \tnodes$ such that $\set{u_1, \ldots, u_{\arityof{\arel}}} \subseteq
  \alabel(n)$, called the \emph{witness} of $\tuple{u_1, \ldots,
    u_{\arityof{\arel}}} \in \struc(\arel)$,
\item\label{it2:reduced-td} every leaf witnesses exactly one tuple $\tuple{u_1, \ldots,
  u_{\arityof{\arel}}} \in \struc(\arel)$,
\item\label{it3:reduced-td} $\tree$ is a binary tree i.e., a tree where each node has at
  most two children,
\item\label{it4:reduced-td} if $n \in \tnodes$ has two children $m_1,m_2 \in \tnodes$
  then $\alabel(n) = \alabel(m_1) = \alabel(m_2)$,
\item\label{it5:reduced-td} if $n \in \tnodes$ has one child $m \in \tnodes$ then either
  $\alabel(n) = \alabel(m)$ and $m$ witnesses a tuple $\tuple{u_1,
  \ldots, u_{\arityof{\arel}}} \in \struc(\arel)$, or
  $\cardof{\alabel(n) \setminus \alabel(m)} = \cardof{\alabel(m)
  \setminus \alabel(n)} = 1$,
\item\label{it6:reduced-td} $\cardof{\alabel(n)} = k+1$, for all $n \in \tnodes$.
\end{compactenum}
\end{definition}
\end{myTextE}
\begin{myLemmaE}\label{lemma:reduced-tree-decompositions}
If a structure $\struc$ has a tree decomposition of width $k$, then it
also has a reduced tree decomposition of width $k$.
\end{myLemmaE}
\begin{proofE}
  The properties (\ref{it1:reduced-td}-\ref{it6:reduced-td}) can
  be proven directly from Definition~\ref{def:treewidth}. The main
  tool is the introduction of intermediate nodes into the tree
  decomposition. We list some of the cases. For
  (\ref{it1:reduced-td}), if a node witnesses more than one tuple,
  then we introduce an intermediate node between itself and its parent
  that is labeled by the same set of vertices. The intermediate node
  then becomes the witness for the this tuple. This process can be
  iterated until every node witnesses at most one tuple.  For
  (\ref{it3:reduced-td}), if a node $n_0$ has children
  $m_0,\cdots,m_l$ we can introduce new nodes $n_1,\ldots,n_{l-1}$
  with $\alabel(m_0) = \alabel(m_i)$, and create a new tree
  decomposition which agrees with the old tree decomposition except
  that we remove the edges from $n_0$ to the children $m_0,\cdots,m_l$
  and add edges from $n_i$ to $m_i$ and $n_{i+1}$ for all $i<l-1$ and
  edges from $n_{l-1}$ to $m_{l-1}$ and $m_l$.  For
  (\ref{it6:reduced-td}), note that we always consider structures over
  some infinite universe.  Hence, we can always extend the labels of
  the tree-decomposition with some fresh vertices from the
  universe.
\end{proofE}

\begin{myLemmaE}
\label{lem:sid-implies-tree-decomposition}
Let $\struc \in \strucof{\signature \cup \set{\domsymb}}$ be a
structure and $\store$ be a store, such that $(\struc,\store)
\models_{\twsid{k}} \apred(x_1,\ldots,x_{k+1})$ and $\store(x_i)
\not\in \struc(\domsymb)$, for all $1 \le i \le k+1$. Then, there
exists a reduced tree decomposition $\tree =
(\tnodes,\tedges,r,\alabel)$ of $\struc$, such that $\twof{\tree}=k$,
$\alabel(r)=\set{\store(x_1),\ldots,\store(x_{k+1})}$ and $\alabel(n)
\subseteq \struc(\domsymb) \cup
\{\store(x_1),\ldots,\store(x_{k+1})\}$, for all $n \in \tnodes$.
\end{myLemmaE}
\begin{proofE}
  We prove the claim by induction on the number of rule
  applications. The claim clearly holds for the base case, by rule
  (\ref{rule:gen-rel}). We now consider the rule (\ref{rule:gen-comp})
  i.e., we assume that
  $(\struc,\store) \models_{\twsid{k}} \apred(x_1,\ldots,x_{k+1})
  * \apred(x_1,\ldots,x_{k+1})$. Then, there exist structures
  $\struc_1$ and $\struc_2$, such that
  $(\struc_i,\store) \models_{\twsid{k}} \apred(x_1,\ldots,x_{k+1})$,
  for $i = 1,2$ and $\struc_1 \comp \struc_2 = \struc$. We note that
  the latter implies that $\struc_1(\domsymb) \cap \struc_2(\domsymb)
  = \emptyset$ ($\dagger$). We now apply the inductive hypothesis and
  obtain reduced tree decompositions $\tree_i$ for $\struc_i$ whose
  respective roots are labelled by
  $\{\store(x_1),\ldots,\store(x_{k+1})\}$ and
  $\alabel(n) \subseteq \struc(\domsymb) \cup \{\store(x_1),\ldots,\store(x_{k+1})\}$,
  for all nodes $n$ of $\tree_i$ and all $i = 1,2$. We obtain a
  reduced tree decomposition for $\struc$ by composing $\tree_1$ and
  $\tree_2$ with a fresh root node labelled by
  $\{\store(x_1),\ldots,\store(x_{k+1})\}$. Note that $\tree$ is
  indeed a tree decomposition because the only elements that may
  appear in labels of both $\tree_1$ and $\tree_2$ must belong to
  $\{\store(x_1),\ldots,\store(x_{k+1})\}$, by ($\dagger$). Clearly,
  we have
  $\alabel(n) \subseteq \struc_1(\domsymb) \cup \struc_2(\domsymb) \cup \{\store(x_1),\ldots,\store(x_{k+1})\}
  = \struc(\domsymb) \cup \{\store(x_1),\ldots,\store(x_{k+1})\}$, for
  all nodes $n$ of the resulting tree decomposition $\tree$. We now
  consider the rule (\ref{rule:gen-exists}) i.e., we assume that
  $(\struc,\store) \models_{\twsid{k}} \exists y ~.~ \domsymb(y)
  * \apred(x_1, \ldots,x_{k+1})[x_i/y]$. Then, there is an element
  $\vertex \not\in \set{\store(x_1), \ldots, \store(x_{k+1})}$ such
  that
  $(\struc',\store[x_i \leftarrow \vertex]) \models_{\twsid{k}} \apred(x_1, \ldots,x_{k+1})$,
  where the structure $\struc'$ agrees with $\struc$, except that
  $\struc(\domsymb) = \struc'(\domsymb) \setminus \set{\vertex}$.  We
  now apply the inductive hypothesis and obtain a reduced tree
  decomposition $\tree_1$ for $\struc'$ whose root is labelled by
  $(\{\store(x_1),\ldots,\store(x_{k+1})\} \setminus \{\store(x_i)\}) \cup \{\vertex\}$
  and $\alabel(n) \subseteq
  ((\struc'(\domsymb) \cup \{\store(x_1),\ldots,\store(x_{k+1})\}) \setminus \{\store(x_i)\}) \cup \{\vertex\}$,
  for all nodes $n$ of $\tree_1$ ($\ddagger$).  We can now obtain a
  reduced tree decomposition $\tree$ for $\struc$ by composing
  $\tree_1$ with an additional root labelled by
  $\{\store(x_1),\ldots,\store(x_{k+1})\}$.  Note that $\tree$ is
  indeed a tree decomposition because $\store(x_i) \not\in \alabel(n)$
  for every node $n$ of $\tree_1$, because of ($\ddagger$),
  $\vertex \neq \store(x_i)$ and the assumption that
  $\store(x_i) \not\in \struc(\domsymb)$.  We have
  $\alabel(n) \subseteq \struc'(\domsymb) \cup \{\store(x_1),\ldots,\store(x_{k+1})\} \cup \{\vertex\}
  = \struc(\domsymb) \cup \{\store(x_1),\ldots,\store(x_{k+1})\}$ for
  all nodes $n$ of the resulting tree decomposition $\tree$.
\end{proofE}

\begin{myTextE}
  \begin{definition}\label{def:struc-t}
    If $\tree=(\tnodes,\tedges,r,\alabel)$ is a tree decomposition of
    a structure $\struc \in \strucof{\signature}$, we denote by
    $\plustd{\tree}{\struc} \in \strucof{\signature \cup
      \set{\domsymb}}$ the structure that agrees with $\struc$ over
    $\signature$, such that, moreover,
    $(\plustd{\tree}{\struc})(\domsymb) = \left(\bigcup_{n \in
      \tnodes} \alabel(n)\right) \setminus \alabel(r)$.
  \end{definition}
\end{myTextE}

\begin{myLemmaE}\label{lem:tree-decomposition-implies-sid}
  Let $\struc$ be a structure with $\twof{\struc} \le k$ witnessed by
  some reduced tree decomposition $\tree=(\tnodes,\tedges,r,\alabel)$,
  such that $\alabel(r)=\{\vertex_1,\ldots,\vertex_{k+1}\}$ and let
  $\store$ be a store with $\store(x_i) = \vertex_i$, for all
  $i \in \interv{1}{k+1}$. Then, we have
  $(\plustd{\tree}{\struc},\store) \models_{\twsid{k}} \apred(x_1, \ldots,
  x_{k+1})$.
\end{myLemmaE}
\begin{proofE}
  The proof goes by induction on the structure of $\tree$.  The claim
  clearly holds for the base case, where $\tree$ consists of a single
  leaf, by rule (\ref{rule:gen-rel}). Consider first the case where
  the root of $\tree$ has two children. The subtrees $\tree_1$ and
  $\tree_2$ rooted in the two children induce substructures $\struc_1$
  and $\struc_2$ of $\struc$, where $\tuple{v_1, \ldots,
  v_{\arityof{\arel}}} \in \struc_i(\arel)$ iff $\tuple{v_1, \ldots,
  v_{\arityof{\arel}}}$ is witnessed by some leaf of $\tree_i$, for $i
  = 1,2$. Because $\tree$ is a reduced tree decomposition, there is at
  most one leaf that witnesses a tuple $\tuple{v_1, \ldots,
  v_{\arityof{\arel}}} \in \struc_i(\arel)$.  Hence, we have $\struc
  = \struc_1 \comp \struc_2$. Because the only elements that can
  appear as labels in both $\tree_1$ and $\tree_2$ are
  $\{\vertex_1,\ldots,\vertex_{k+1}\}$, we also get that
  $\plustd{\tree}{\struc}
  = \plustd{\tree_1}{\struc_1} \comp \plustd{\tree_2}{\struc_2}$. From
  the inductive hypothesis we obtain
  $(\plustd{\tree_i}{\struc_i},\store) \models_{\twsid{k}} \apred(x_1, \ldots,
  x_{k+1})$, for $i = 1,2$.  Hence
  $(\plustd{\tree_1}{\struc_1} \comp \plustd{\tree_2}{\struc_2}, \store) \models_{\twsid{k}} \apred(x_1, \ldots,
  x_{k+1}) * \apred(x_1, \ldots, x_{k+1})$, thus
  $(\plustd{\tree}{\struc},\store) \models_{\twsid{k}} \apred(x_1, \ldots,
  x_{k+1})$, by rule (\ref{rule:gen-comp}). Consider now the case
  where the root of $\tree$ has a single child which is not a leaf and
  consider the subtree $\tree_1$ rooted at this child. Then, there is
  an element $\vertex \not\in \set{\vertex_1, \ldots, \vertex_{k+1}}$,
  such that the root of $\tree_1$ is labeled by
  $\{\vertex_1,\ldots,\vertex_{i-1},\vertex,\vertex_{i+1},\ldots,\vertex_{k+1}\}$.
  By the inductive hypothesis, we obtain
  $(\plustd{\tree_1}{\struc},\store[x_i \leftarrow \vertex]) \models_{\twsid{k}} \apred(x_1, \ldots,
  x_{k+1})$. We recognize that $\plustd{\tree_1}{\struc}$ agrees with
  $\plustd{\tree}{\struc}$ except that $\vertex \in
  (\plustd{\tree}{\struc})(\domsymb)$.  Hence,
  $(\plustd{\tree}{\struc},\store) \models_{\twsid{k}} \exists y
  ~.~ \domsymb(y) * \apred(x_1,\ldots,x_{k+1})[x_i / y]$, thus
  $(\plustd{\tree}{\struc},\store) \models_{\twsid{k}} \apred(x_1,\ldots,x_{k+1})$,
  by rule (\ref{rule:gen-exists}).
\end{proofE}

\begin{lemmaE}\label{lemma:k-sid}
  For any structure $\struc \in \strucof{\signature}$, we have
  $\twof{\struc} \leq k$ if and only if there is a
  $\domsymb$-extension $\struc'$ of $\struc$ with
  $\struc' \models_{\twsid{k}} \apred_k()$.
\end{lemmaE}
\begin{proofE}
``$\Rightarrow$'' If $\struc$ has tree decomposition of width at most
$k$, then it also has a reduced tree decomposition
$\tree=(\tnodes,\tedges,r,\alabel)$ of width $k$, by
Lemma~\ref{lemma:reduced-tree-decompositions}.  Let
$\alabel(r)=\set{\vertex_1, \ldots, \vertex_{k+1}}$ and $\store$ be a
store such that $\store(x_i) = \vertex_i$, for all
$i \in \interv{1}{k+1}$.  By
Lemma~\ref{lem:tree-decomposition-implies-sid},
$(\plustd{\tree}{\struc},\store) \models_{\twsid{k}} \apred(x_1, \ldots,
x_{k+1})$.  By adding $\set{\vertex_1, \ldots, \vertex_{k+1}}$ to the
interpretation of $\domsymb$ in $\plustd{\tree}{\struc}$, we obtain a
$\domsymb$-extension $\struc'$ of $\struc$ that satisfies
$\struc' \models_{\twsid{k}} \apred_k()$, by rule
(\ref{rule:gen-top}).

  ``$\Leftarrow$'' Let $\struc'$ be a $\domsymb$-extension of $\struc$
  with $\struc' \models_{\twsid{k}} \apred_k()$.  By rule
  (\ref{rule:gen-top}), there exists a $\domsymb$-extension $\struc''$
  of $\struc$ and a store $\store$, with
  $\store(x_i) \neq \store(x_j)$, for all $i \neq
  j \in \interv{1}{k+1}$ and
  $\store(x_1), \ldots \store(x_{k+1}) \not\in \struc''(\domsymb)$,
  such that $(\struc'',\store) \models_{\twsid{k}} \apred(x_1, \ldots,
  x_{k+1})$.
  By Lemma \ref{lem:sid-implies-tree-decomposition}, there
  exists a reduced tree decomposition $\tree$ of $\struc''$ of width
  $k$. Thus $\twof{\struc} \leq k$, because $\struc$ is a substructure
  of $\struc''$.
\end{proofE}


The second step of our construction is the annotation of the rules in
$\twsid{k}$ with $\qrof{\phi}$-types, in order to obtain an SID
$\twformsid{k}{\phi}$ describing the models of an \mso\ sentence
$\phi$, of treewidth at most $k$.
For this, we consider the set of
ports $\ports = \set{\acst_{\ncst+1}, \ldots, \acst_{\ncst+k+1}}$
disjoint from $\signature$, used to encode the values of the variables
$x_1,\ldots,x_{k+1}$:

  \begin{definition}\label{def:encode} Let $\signature
    = \set{\arel_1, \ldots, \arel_\nrel, \acst_1, \ldots,
      \acst_\ncst}$ be a signature and $\ports = \set{\acst_{\ncst+1},
      \ldots, \acst_{\ncst+k+1}}$ be constants not in
    $\signature$. For a structure $\struc \in \strucof{\signature}$
    and a store $\store : \set{x_1, \ldots, x_m} \rightarrow
    \universe$, we denote by $\encof{k+1}{\struc}{\store} \in
    \strucof{\signature \cup \set{\acst_{\ncst+1}, \ldots,
        \acst_{\ncst+k+1}}}$ the structure that agrees with $\struc$
    over $\signature$ and maps $\acst_{\ncst+i}$ to $\store(x_i)$, for
    all $i \in \interv{1}{k+1}$.
\end{definition}

The correctness of our construction will rely on the fact that the disjoint union behaves like glueing for two structures that only overlap at the ports:

\begin{lemmaE}\label{lemma:abstract-glue}
  Given an integer $r\geq0$, a store $\store$ and disjoint compatible
  structures $\struc_1,\struc_2 \in \strucof{\signature}$ with
  $\Rel{\struc_1} \cap \Rel{\struc_2} \subseteq \{\store(x_1),\ldots,\store(x_{k+1})\}$,
  we have $\typeof{r}{\encof{k+1}{\struc_1 \comp \struc_2}{\store}}
  = \absglueof{\typeof{r}{\encof{k+1}{\struc_1}{\store}}}{\typeof{r}{\encof{k+1}{\struc_2}{\store}}}$.
\end{lemmaE}
\begin{proofE}
  Let us consider $\struc_1' = \encof{k+1}{\struc_1}{\store}$ and
  $\struc_2' = \encof{k+1}{\struc_2}{\store}$.  In order to apply
  glueing we will now consider two standard structures isomorphic to
  $\struc_1'$ and $\struc_2'$, respectively.  We note that
  $\Dom{\struc_1'} \cap \Dom{\struc_2'} \subseteq \{\struc_2'(\acst_{\ncst_1}),\ldots,\struc_2'(\acst_{\ncst+k+1})\}$
  because of our assumption that
  $\Rel{\struc_1} \cap \Rel{\struc_2} \subseteq \{\store(x_1),\ldots,\store(x_{k+1})\}$).
  We further note that $\struc_1'(\acst_{\ncst_i})
  = \struc_2'(\acst_{\ncst_i})$, for all
  $i \in \interv{1}{\ncst+k+1}$, because $\struc_1$ and $\struc_2$ are
  compatible and the interpretation of the additional constants
  $\acst_{\ncst+1},\ldots,\acst_{\ncst+k+1}$ has been chosen
  w.r.t. the same store $\store$. Hence, we can choose some
  partitioning $\mathcal{U}_1 \uplus \mathcal{U}_2 = \universe$ such
  that $\mathcal{U}_1$ and $\mathcal{U}_2$ are countably infinite,
  $\Dom{\struc_1'} \subseteq \mathcal{U}_1$ and
  $\Dom{\struc_2'} \setminus \{\struc_2'(\acst_1),\ldots,\struc_2'(\acst_{\ncst+k+1})\} \subseteq \mathcal{U}_2$.
  We can now choose a structure $\struc''_2$ with
  $\Dom{\struc''_2} \subseteq \mathcal{U}_2$ that is isomorphic to
  $\struc'_2$ and that agrees with $\struc'_2$ except for
  $\acst_1,\ldots\acst_{\ncst+k+1}$, whose interpretation is chosen as
  $\struc''_2(\acst_1), \ldots, \struc''_2(\acst_{\ncst+k+1}) \in \mathcal{U}_2 \setminus
  (\Dom{\struc_1'} \cup \Dom{\struc_2'})$.  Then,
  $(\mathcal{U}_1,\struc_1')$ [resp. $(\mathcal{U}_2,\struc''_2)$] is
  isomorphic to $(\universe,\struc_1)$
  [resp. $(\universe,\struc_2)$]. In particular, they have the same
  type i.e., $\typeof{r}{\mathcal{U}_1,\struc_1'}
  = \typeof{r}{\universe,\struc_1'}$ and
  $\typeof{r}{\mathcal{U}_2,\struc''_2}
  = \typeof{r}{\universe,\struc_2'}$.  Moreover, we have
  $\glueof{(\mathcal{U}_1,\struc_1')}{(\mathcal{U}_2,\struc''_2)}
  = \encof{k+1}{\struc_1 \comp \struc_2}{\store}$.  We compute:
  \vspace*{-.5\baselineskip}
  \[\begin{array}{l} \typeof{r}{\encof{k+1}{\struc_1 \comp \struc_2}{\store}}
  = \\ \typeof{r}{\glueof{(\mathcal{U}_1,\struc_1')}{(\mathcal{U}_2,\struc''_2)}}
  = \\ \absglueof{\typeof{r}{(\mathcal{U}_1,\struc_1')}}{\typeof{r}{(\mathcal{U}_2,\struc''_2)}}
  = \\ \absglueof{\typeof{r}{(\universe,\struc_1')}}{\typeof{r}{(\universe,\struc_2')}}
  = \\ \absglueof{\typeof{r}{\encof{k+1}{\struc_1}{\store}}}{\typeof{r}{\encof{k+1}{\struc_2}{\store}}} \end{array}\]
\end{proofE}

\begin{myLemmaE}\label{lem:refined-SID-is-sound}
Let $k\geq1$ be an integer, $\phi$ be an \mso\ sentence,
$\struc \in \strucof{\signature\cup\set{\domsymb}}$ be a structure and
$\store$ be a store such that $(\struc,\store)
\models_{\twformsid{k}{\phi}} \apred^\atype(x_1, \ldots, x_{k+1})$,
$\store(x_i) \not\in \struc(\domsymb)$ for all $i \in
\interv{1}{k+1}$, and $\store(x_i) \neq \store(x_j)$ for all $i \neq j$.
Then,
$\typeof{\qrof{\phi}}{\encof{k+1}{\minusdom(\struc)}{\store}} =
\atype$, where $\minusdom(\struc) \in \strucof{\signature}$ is the
restriction of $\struc$ to $\signature$.
\end{myLemmaE}
\begin{proofE}
  We prove the claim by induction on the number of rule applications.
  The claim clearly holds for the base case, by a rule of type
  (\ref{rule:ref-rel}). We now consider a rule of type
  (\ref{rule:ref-comp}), i.e., we have $(\struc,\store)
  \models_{\twformsid{k}{\phi}} \apred^{\atype_1}(x_1, \ldots,
  x_{k+1}) * \apred^{\atype_2}(x_1, \ldots, x_{k+1})$, such that
  $\atype = \absglueof{\atype_1}{\atype_2}$. Then, there are
  structures $\struc_1$ and $\struc_2$ with $(\struc_i, \store)
  \models_{\twformsid{k}{\phi}} \apred^{\atype_i}(x_1, \ldots,
  x_{k+1})$, for $i = 1,2$ and $\struc_1 \comp \struc_2 = \struc$.
  By the
  inductive hypothesis, we have that
  $\typeof{\qrof{\phi}}{\encof{k+1}{\minusdom(\struc_i)}{\store}}
  = \atype_i$, for $i=1,2$.  Because every derivation of
  $\twformsid{k}{\phi}$ is also a derivation of $\twsid{k}$,
  obtained by removing the type annotations from the rules in
  $\twformsid{k}{\phi}$, we get that
  $(\struc_i,\store) \models_{\twsid{k}} \apred(x_1, \ldots,
  x_{k+1})$, for $i=1,2$.  By
  Lemma~\ref{lem:sid-implies-tree-decomposition} we have that
  $\Rel{\struc_i} \subseteq \struc_i(\domsymb) \cup \{\store(x_1), \ldots, \store(x_{k+1})\}$,
  for $i=1,2$. Since $\struc = \struc_1 \comp \struc_2$, we have that
  $\struc_1(\domsymb) \cap \struc_2(\domsymb) = \emptyset$ and
  $\minusdom(\encof{k+1}{\struc}{\store})
  = \minusdom(\encof{k+1}{\struc_1}{\store}) \comp \minusdom(\encof{k+1}{\struc_2}{\store})$.
  We compute
  \vspace*{-.5\baselineskip}
  \[\begin{array}{l} \typeof{\qrof{\phi}}{\encof{k+1}{\minusdom(\struc)}{\store}}
  = \\ \absglueof{\typeof{\qrof{\phi}}{\encof{k+1}{\minusdom(\struc_1)}{\store}}}{\typeof{\qrof{\phi}}{\encof{k+1}{\minusdom(\struc_2)}{\store}}}
  = \\ \absglueof{\atype_1}{\atype_2} = \atype \end{array}\]

  \vspace*{-.5\baselineskip}\noindent
  by Lemma~\ref{lemma:abstract-glue}.  We now consider rules of type
  (\ref{rule:ref-exists}) i.e., we have $(\struc,\store)
  \models_{\twformsid{k}{\phi}} \exists y ~.~ \domsymb(y) *
  \apred^{\atype_1}(x_1, \ldots,x_{k+1})[x_i/y]$, for some $i \in
  \interv{1}{k+1}$, such that $\atype =
  \absglueof{\absfgcst{\ncst+i}(\atype_1)}{\rho_i}$ for the type
  $\rho_i$ of a standard structure $\astruc \in
  \strucof{\set{\acst_{\ncst+i}}}$ with a singleton universe, for $i
  \in \interv{1}{k+1}$. Then, there is an element $\vertex \in
  \universe$, such that $(\struc',\store[x_i \leftarrow \vertex])
  \models_{\twformsid{k}{\phi}} \apred^{\atype_1}(x_1,
  \ldots,x_{k+1})$, where the structure $\struc'$ agrees with
  $\struc$, except that $\domsymb$ does not hold for $\vertex$ in
  $\struc'$.  By the inductive hypothesis,
  $\typeof{\qrof{\phi}}{\encof{k+1}{\minusdom(\struc')}{\store[x_i
        \leftarrow \vertex]}} = \atype_1$.  Because every derivation
  of $\twformsid{k}{\phi}$ is also a derivation of
  $\twsid{k}$, obtained by removing the type annotations
  from the rules in $\twformsid{k}{\phi}$, we get that
  $(\struc_i,\store) \models_{\twsid{k}} \apred(x_1, \ldots,
  x_{k+1})$, for $i=1,2$.  By
  Lemma~\ref{lem:sid-implies-tree-decomposition} we have that
  $\Rel{\struc'} \subseteq \struc'(\domsymb) \cup \{\store(x_1),
  \ldots, \store(x_{k+1})\} \setminus \{\store(x_i)\} \cup \{u\}$.
  Because of $\store(x_i) \neq u$ (due to the assumption $\store(x_i)
  \not\in \struc(\domsymb)$) and because of $\store(x_i) \neq
  \store(x_j)$ for all $i \neq j$, we get that
  $\encof{k+1}{\minusdom(\struc)}{\store} =
  \glueof{\fgcst{\ncst+i}(\encof{k+1}{\minusdom(\struc')}{\store[x_i
        \leftarrow \vertex]})}{\astruc'}$, where $\astruc' \in
  \strucof{\set{\acst_{\ncst+i}}}$ is the standard structure with
  singleton universe $\{\store(x_i)\}$.  Because $\astruc$ is
  isomorphic to $\astruc'$, we obtain:
  \vspace*{-.5\baselineskip}
  \[\begin{array}{l} \typeof{\qrof{\phi}}{\encof{k+1}{\minusdom(\struc)}{\store}} = \\
  \typeof{\qrof{\phi}}{ \glueof{\fgcst{\ncst+i}(\encof{k+1}{\minusdom(\struc')}{\store[x_i \leftarrow \vertex]})}{\astruc'}} = \\
  \absglueof{\typeof{\qrof{\phi}}{\fgcst{\ncst+i}(\encof{k+1}{\minusdom(\struc')}{\store[x_i \leftarrow \vertex]})}}{\typeof{\qrof{\phi}}{\astruc'}} = \\
  \absglueof{\absfgcst{\ncst+i}(\typeof{\qrof{\phi}}{\encof{k+1}{\minusdom(\struc')}{\store[x_i \leftarrow \vertex]}})}{\typeof{\qrof{\phi}}{\astruc}} = \\
  \absglueof{\absfgcst{\ncst+i}(\atype_1)}{\rho_i} = \atype \end{array}\]
\end{proofE}

\begin{myLemmaE}\label{lem:refined-SID-is-complete}
Let $k\geq1$ be an integer, $\phi$ be an \mso\ sentence, $\struc \in
\strucof{\signature}$ be a structure of treewidth $\twof{\struc} \le
k$, witnessed by some reduced tree decomposition
$\tree=(\tnodes,\tedges,r,\alabel)$, and $\store$ be a store with $\alabel(r)=\{\store(x_1),\ldots,\store(k+1)\}$ and $\store(x_i) \neq \store(x_j)$ for all $i \neq j$, such that
$\typeof{\qrof{\phi}}{\encof{k+1}{\struc}{\store}} = \atype$ and $\phi
\in \atype$.
Then, we have $(\plustd{\tree}{\struc},\store)
\models_{\twformsid{k}{\phi}} \apred^\atype(x_1, \ldots, x_{k+1})$.
\end{myLemmaE}
\begin{proofE}
  The proof proceeds by induction on the structure of $\tree$.
  Clearly the claim holds for the base case, by a rule of type
  (\ref{rule:ref-rel}). For the inductive step, we assume first that
  the root of $\tree$ has two children. The subtrees $\tree_1$ and
  $\tree_2$ rooted in the two children induce substructures $\struc_1$
  and $\struc_2$ of $\struc$, where $\tuple{v_1, \ldots,
    v_{\arityof{\arel}}} \in \struc_i(\arel)$ if and only if
  $\tuple{v_1, \ldots, v_{\arityof{\arel}}}$ is witnessed by some leaf
  of the respective subtree.  Because $\tree$ is a reduced tree
  decomposition, there is at most one leaf that witnesses a tuple
  $\tuple{v_1, \ldots, v_{\arityof{\arel}}} \in \struc_i(\arel)$.
  Hence, we have $\struc = \struc_1 \comp \struc_2$ and
  $\encof{k+1}{\struc}{\store} = \encof{k+1}{\struc_1}{\store} \comp
  \encof{k+1}{\struc_2}{\store}$.  Because the only elements that can
  appear as labels in $\tree_1$ and $\tree_2$ are
  $\store(x_1),\ldots,\store(x_{k+1})$ (as these are the labels of the root of the tree decomposition), we get
  that $\Rel{\struc_1} \cap \Rel{\struc_2} \subseteq
  \set{\store(x_1),\ldots,\store(x_{k+1})}$.  Let $\atype_i =
  \typeof{\qrof{\phi}}{\encof{k+1}{\struc_i}{\store}}$, for $i=1,2$.
  From the inductive hypothesis, we obtain
  $(\plustd{\tree_i}{\struc_i},\store) \models_{\twformsid{k}{\phi}}
  \apred^{\atype_i}(x_1, \ldots, x_{k+1})$, for $i=1,2$. Let $\atype
  \isdef \absglueof{\atype_1}{\atype_2}$.  By
  Lemma~\ref{lemma:abstract-glue}, we obtain:
  \vspace*{-.5\baselineskip}
  \[\begin{array}{l}
  \typeof{\qrof{\phi}}{\encof{k+1}{\struc}{\store}} = \typeof{\qrof{\phi}}{\encof{k+1}{\struc_1}{\store} \comp \encof{k+1}{\struc_2}{\store}} = \\
   \absglueof{\typeof{\qrof{\phi}}{\encof{k+1}{\struc_1}{\store}}}{\typeof{\qrof{\phi}}{\encof{k+1}{\struc_2}{\store}}} = \absglueof{\atype_1}{\atype_2}  = \atype
  \end{array}\]

  \vspace*{-.5\baselineskip}\noindent
  Hence, we now get that $(\plustd{\tree}{\struc},\store)
  \models_{\twformsid{k}{\phi}} \apred^\atype(x_1, \ldots, x_{k+1})$,
  by a rule of type (\ref{rule:ref-comp}). We now assume that the root
  of $\tree$ has a single child which is not a leaf. We consider the
  subtree $\tree_1$ rooted at this child. Then, there is an element
  $\vertex \not\in \set{\store(x_1), \ldots, \store(x_{k+1})}$, such
  that the root of $\tree_1$ is labeled by $\set{\store(x_1), \ldots,    \store(x_{i-1}), \vertex,\store(x_{i+1}), \ldots, \store(x_{k+1})}$.
  Let $\atype_1 \isdef \typeof{\qrof{\phi}}{\encof{k+1}{\struc}{\store[x_i \leftarrow \vertex]}}$. By the inductive hypothesis, we obtain that
  $(\plustd{\tree_1}{\struc},\store[x_i \leftarrow \vertex])
  \models_{\twformsid{k}{\phi}} \apred^{\atype_1}(x_1, \ldots,
  x_{k+1})$.
  Let $\atype \isdef \absglueof{\absfgcst{\ncst+i}(\atype_1)}{\rho_i}$ for the type $\rho_i$ of a structure $\astruc \in \strucof{\set{\acst_{\ncst+i}}}$ with singleton universe $\{\store(x_i\}$.
  Because of $\store(x_i) \neq u$ and because of $\store(x_i) \neq \store(x_j)$ for all $i \neq j$,
  we get that $\encof{k+1}{\minusdom(\struc)}{\store} = \glueof{\fgcst{\ncst+i}(\encof{k+1}{\minusdom(\struc')}{\store[x_i \leftarrow \vertex]})}{\astruc}$.
  We compute:
  \vspace*{-.5\baselineskip}
  \[\begin{array}{l}
  \typeof{\qrof{\phi}}{\encof{k+1}{\minusdom(\struc)}{\store}} = \\
  \typeof{\qrof{\phi}}{ \glueof{\fgcst{\ncst+i}(\encof{k+1}{\minusdom(\struc')}{\store[x_i \leftarrow \vertex]})}{\astruc}} = \\
  \absglueof{\typeof{\qrof{\phi}}{\fgcst{\ncst+i}(\encof{k+1}{\minusdom(\struc')}{\store[x_i \leftarrow \vertex]})}}{\typeof{\qrof{\phi}}{\astruc}} = \\
  \absglueof{\absfgcst{\ncst+i}(\typeof{\qrof{\phi}}{\encof{k+1}{\minusdom(\struc')}{\store[x_i \leftarrow \vertex]}})}{\typeof{\qrof{\phi}}{\astruc}} = \\
  \absglueof{\absfgcst{\ncst+i}(\atype_1)}{\rho_i} = \atype
  \end{array}\]

  \vspace*{-.5\baselineskip}\noindent
  We now recognize that $\plustd{\tree_1}{\struc}$ agrees with
  $\plustd{\tree}{\struc}$ except that
  $\plustd{\tree}{\struc}(\domsymb) =
  \plustd{\tree_1}{\struc}(\domsymb) \cup \set{\vertex}$. Hence, we get
  that $(\plustd{\tree}{\struc},\store) \models_{\twformsid{k}{\phi}}
  \exists y ~.~ \domsymb(y) * \apred_\tau(x_1, \ldots,x_{k+1})[x_i/y]$
  i.e., that $(\plustd{\tree}{\struc},\store)
  \models_{\twformsid{k}{\phi}} \apred^\atype(x_1, \ldots, x_{k+1})$
  by a rule of type (\ref{rule:ref-exists}).
\end{proofE}

Then $\twformsid{k}{\phi}$ is the set of rules
from Figure \ref{fig:sids} (b), whose property is stated below:

\begin{propositionE}\label{prop:kform-sid}
  Given $k\geq1$ and an \mso\ sentence $\phi$, for any structure
  $\struc \in \strucof{\signature}$, the following are
  equivalent: \begin{inparaenum}[(1)]
  \item\label{it2:kform-sid} $\struc \Models \phi$ and $\twof{\struc}
    \leq k$, and
  \item\label{it1:kform-sid} there is a $\domsymb$-extension $\struc'$ of $\struc$ with $\struc' \models_{\twformsid{k}{\phi}} \apred_{k,\phi}()$.
  \end{inparaenum}
\end{propositionE}
\begin{proofE}
  ``(\ref{it2:kform-sid}) $\Rightarrow$ (\ref{it1:kform-sid})'' Since
  $\twof{\struc} \leq k$, there exists a reduced tree decomposition
  $\tree=(\tnodes,\tedges,r,\alabel)$ of width $k$, by
  Lemma~\ref{lemma:reduced-tree-decompositions}.  Let $\store$ be a
  store such that $\alabel(r)
  = \set{\store(x_1), \ldots, \store(x_{k+1})}$ and
  $\store(x_i) \neq \store(x_j)$ for all $i \neq
  j \in \interv{1}{k+1}$.  Thus, we obtain
  $(\plustd{\tree}{\struc},\store) \models_{\twformsid{k}{\phi}} \apred^\atype(x_1,\ldots,x_{k+1})$,
  by Lemma~\ref{lem:refined-SID-is-complete}, for
  $\atype \isdef \typeof{\qrof{\phi}}{\encof{k+1}{\struc}{\store}}$.
  Since, moreover, we have assumed that $\struc \Models \phi$, we have
  $\phi \in \atype \isdef \typeof{\qrof{\phi}}{\encof{k+1}{\struc}{\store}}$.
  Let $\struc'$ be the $\domsymb$-extension obtained from
  $\plustd{\tree}{\struc}$ by setting $\struc'(\domsymb)
  = \plustd{\tree}{\struc}(\domsymb) \cup \{\store(x_1), \ldots, \store(x_{k+1})\}$. Then,
  we obtain $\struc' \models_{\twformsid{k}{\phi}} \apred_{k,\phi}()$,
  by a rule of type (\ref{rule:ref-top}).

  ``(\ref{it1:kform-sid}) $\Rightarrow$ (\ref{it2:kform-sid})'' Let
  $\struc'$ be a $\domsymb$-extension of $\struc$ with
  $\struc' \models_{\twformsid{k}{\phi}} \apred_{k,\phi}()$.  Since
  $\struc' \models_{\twformsid{k}{\phi}} \apred_{k,\phi}()$, there
  exists a $\domsymb$-extension $\struc''$ of $\struc$ and a store
  $\store$ with $\store(x_i) \neq \store(x_j)$ for all $i \neq
  j \in \interv{1}{k+1}$, such that
  $(\struc'',\store) \models_{\twformsid{k}{\phi}} \apred^\atype(x_1, \ldots,
  x_{k+1})$, for some type $\atype$, such that $\phi \in \atype$, by a
  rule of type (\ref{rule:ref-top}).  By
  Lemma~\ref{lem:refined-SID-is-sound}, we obtain
  $\typeof{\qrof{\phi}}{\encof{k+1}{\struc}{\store'}} = \atype$, thus
  $\encof{k+1}{\struc}{\store'} \Models \phi$, leading to
  $\struc \Models \phi$.  Moreover, we have $\twof{\struc} \leq k$, by
  Lemma \ref{lemma:k-sid}, because each derivation of
  $\struc' \models_{\twformsid{k}{\phi}} \apred_{k,\phi}()$
  corresponds to a derivation of
  $\struc' \models_{\twsid{k}} \apred_{k}()$, obtained by removing the
  type annotations from the rules in $\twformsid{k}{\phi}$.
\end{proofE}

The above result shows that \slr\ can define the models
$\struc \in \strucof{\signature}$ of a given \mso\ formula whose
treewidth is bounded by a given integer, up to extending the signature
$\signature$ with a monadic relation symbol $\domsymb$, whose
interpretation subsumes the set $\Rel{\struc}$ of elements that occur
in some tuple from $\struc(\arel)$, for some relation symbol
$\arel \in \signature$. It is unclear, for the moment, how to
prove \mso\ $\subseteq^k$ \slr\ without introducing this gadget.

\paragraph{Effectiveness of the Construction.}
The above construction of the SID $\twformsid{k}{\phi}$ is effective
apart from rule~\ref{rule:ref-rel}, where one needs to determine the
type of a structure $\astruc$ with infinite universe.  However, we
argue in the following that determining this type can be reduced to
computing the type of a finite structure; the type of such a structure
can in turn be determined by solving finitely many \mso\ model
checking problems on finite structures, each of which being known to
be \pspace-complete \cite{DBLP:conf/stoc/Vardi82}.

Given an integer $k\geq0$ and a structure $\astruc = (\universe,\struc) \in
\strucof{\signature}$, we define the finite
structure $\astruc^k = (\adomof{}^k,\struc)$, where $\adomof{}^k
\isdef \Dom{\struc} \cup \set{\vertex_1, \ldots, \vertex_k}$, for
pairwise distinct elements $\vertex_1, \ldots, \vertex_k \in \universe
\setminus \Dom{\struc}$.
Then, for any quantifier rank $r$, the structures $\astruc$ and
$\astruc^{2^r}$ have the same $r$-type, as shown below:

\begin{lemmaE}\label{lemma:single-type}
  Given $r\geq0$ and $\astruc = (\universe,\struc) \in
  \strucof{\signature}$, we have $\typeof{r}{\astruc} = \typeof{r}{\astruc^{2^r}}$.
\end{lemmaE}
\begin{proofE}
  We consider the $r$-round MSO Ehrenfeucht-Fra\"{i}ss\'{e} game: In
  every round, Spoiler picks a vertex $\vertex$ or a set of vertices
  $\vertexSet$ in $\astruc$ or $\astruc^{2^r}$ and Duplicator answers
  with a vertex or a set of vertices in the other structure.
  Duplicator wins iff after $r$-rounds the substructures induced by
  the selected vertices are isomorphic.  We now sketch a winning
  strategy for Duplicator (as the argument is standard, we leave the
  details to the reader): For vertices in $\Dom{\struc}$ the
  Duplicator plays the same vertices in the respective other structure
  (this applies to single vertices as well as sets of vertices).  For
  the vertices that do not belong to $\Dom{\struc}$ the Duplicator
  selects vertices that do not belong to $\Dom{\struc}$ in the other
  structure such that the selected vertices belong to the same subsets
  played in the previous rounds; care has to be taken for the involved
  cardinalities, if playing in $\astruc^{2^r}$ the Duplicator limits
  himself to at most $2^{r-k-1}$ vertices that belong to any
  combination of previously played subsets or the complement of these
  subsets, where $k$ is the number of previously played rounds; this
  choice is sufficient such that after $r$-rounds the substructures
  induced by the selected vertices are isomorphic.
\end{proofE}

\section{The Remaining Cases}
\label{sec:remaining}

We present the results for the problems from Table
\ref{tab:expressiveness}, not already covered in \S\ref{sec:slr-mso},
\S\ref{sec:mso-slr} and \S\ref{sec:slr-so}.

\paragraph{\sol\ $\not\subseteq^{(k)}$ \mso}
First, \sol\ $\not\subseteq^{k}$ \mso\ is a consequence of the fact
that any non-recognizable context-free word language, corresponding to a
non-\mso-definable family of structures can be defined in \slr. Since
\slr\ $\subseteq$\ \sol, we obtain that \sol\ $\not\subseteq^k$
\mso. Moreover, \sol\ $\not\subseteq$ \mso\ follows from the fact that
our counterexample involves only structures of treewidth $1$ (i.e.,
lists encoding words as in \S\ref{sec:slr-mso}).

\paragraph{\slr\ $\subseteq^k$ \sol}
By applying the translation of \slr\ to \sol\ from \S\ref{sec:slr-so}
to $\twsid{k}$ (Fig. \ref{fig:sids}a) and to a given SID $\asid$ defining a predicate $\apred$ of zero arity, 
respectively, and taking the conjunction of the results, we obtain an
\sol\ formula that defines the \slr\ models of $(\apred(),\asid)$ of treewidth at most
$k$, thus proving \slr\ $\subseteq^k$ \sol.

\paragraph{\solmso\ $\subseteq^k$ \solmso}
For each given $k \geq 1$, there exists an \mso\ formula $\theta_k$
that defines the structures of treewidth at most $k$ \cite[Proposition
  5.11]{courcelle_engelfriet_2012}. This is a consequence of the Graph
Minor Theorem proved by Robertson and Seymour
\cite{DowneyFellowsBook}, combined with the fact that bounded
treewidth graphs are closed under taking minors and that the property
of having a given finite minor is \mso-definable\footnote{The
original proof of Robertson and Seymour does not build $\theta_k$
effectively, see
\cite{10.5555/1347082.1347153} for an effective proof.}. Then, for any given
\solmso\ formula $\phi$, the \solmso\ formula $\phi \wedge \theta_k$
defines the models of $\phi$ of treewidth at most $k$.

\paragraph{Open Problems}
The following problems from Table \ref{tab:expressiveness} are
currently open: 
\slr\ $\subseteq^k$\ \slr\ and \sol\ $\subseteq^k$
\slr, both conjectured to have a negative answer.
In particular, the difficulty concerning \slr\ $\subseteq^k$\ \slr\ is
that, in order to ensure treewidth boundedness, it seems necessary to
force the composition of structures to behave like glueing (see the
definition of $\twsid{k}$ in Fig. \ref{fig:sids}a), which is however
difficult to ensure without an additional predicate symbol.

Since \mso\ $\subseteq^k$ \slr\ but \mso\ $\not\subseteq$ \slr, it is
natural to ask for the existence of a fragment of \slr\ that describes
only \mso-definable families of structures of bounded
treewidth. Moreover, in earlier work \cite{DBLP:conf/cade/IosifRS13},
we showed that a significant fragment of \seplog, given by three
restrictions on the syntax of the rules, can be embedded in
\mso\ and defines only structures of bounded treewidth. Unfortunately,
since \slr\ can define context-free languages
(Prop. \ref{prop:cfg-slr}), the
\mso-definability of the set of models of a \slr\ formula is
undecidable, as a consequence of the undecidability of the
recognizability of context-free languages \cite{Greibach68}. On the
other hand, the treewidth-boundedness of the set of models of a \slr\
formula is an open problem, related to the open problem
\slr\ $\subseteq^k$\ \slr\ above.

\section{Conclusions and Future Work}

We have compared the expressiveness of \slr, \mso\ and \sol, in
general and for models of bounded treewidth. Interestingly, we found
that \slr\ and \mso\ are, in general, incomparable and subsumed by
\sol, whereas the models of bounded treewidth of \mso\ can be defined
by \slr, modulo augmenting the signature with a unary relation symbol
used to store the elements that occur in the original
structure.

Future work concerns answering the open problems of the
\slr-definability of structures defined by either an \slr\ or
\sol\ formula, of treewidth bounded by a given integer. A cornerstone
is deciding whether a given SID defines only structures of bounded
treewidth. Answering these questions would help the definition of
\slr\ fragments with a decidable entailment problem, useful for system
verification.

\bibliographystyle{abbrv}
\bibliography{refs}

\end{document}